**Ultrafast laser synthesis of zeolites**

*Sezin Galioglu[1,*,†], Mehdi Hagverdiyev[1,*], Meryem M. Doğan[1], Özgün Yavuz[2], Ü. Seleme Nizam[2], Ghaith Makey[1], Aladin Şura[2], Mesut Laçin[2], Burcu Akata Kurç[4], Parviz Elahi[5], F. Ömer Ilday[2,3], Serim Ilday[2,3,†]*

[1] UNAM-National Nanotechnology Research Center, Bilkent University, Ankara, 06800, Türkiye

[2] Faculty of Electrical Engineering and Information Technology, Ruhr University Bochum, 44801, Germany

[3] Faculty of Physics and Astronomy, Ruhr University Bochum, 44801, Germany

[4] Micro and Nanotechnology Programme, Middle East Technical University, 06800, Ankara, Türkiye

[5] Faculty of Engineering, Özyeğin University, 34794, İstanbul, Türkiye

*The authors have equally contributed to this study.

†To whom correspondence should be addressed: sezin@unam.bilkent.edu.tr & serim.ilday@rub.de

**Abstract**

Research demonstrates that zeolite nucleation and growth can be controlled by fine-tuning chemical composition, temperature, and pressure, resulting in structures with diverse porosities and functionalities. Nevertheless, current energy delivery methods lack the finesse required to operate on the femto- and picosecond timescales of silica polymerization and depolymerization, limiting their ability to direct synthesis with high precision. To overcome this limitation, we introduce an ultrafast laser synthesis technique capable of delivering energy at these timescales with unprecedented spatiotemporal precision. Unlike conventional or emerging approaches, this method bypasses the need for specific temperature and pressure settings, as nucleation and growth are governed by dynamic phenomena arising from nonlinear light-matter interactions, such as convective flows, cavitation bubbles, plasma formation, and shock waves. These processes can be initiated, paused, and resumed within fractions of a second, effectively "freezing" structures at any stage of self-assembly. Using this approach, we trace the entire nucleation and growth pathway of laser-synthesized TPA-silicate-1 zeolites from early oligomer formation to fully developed crystals. The unprecedented spatiotemporal control of this technique unlocks new avenues for manipulating reaction pathways and exploring the vast configurational space of zeolites.

## 1. Introduction

Zeolites are crystalline inorganic materials with well-dispersed and systematically arranged nano and/or micron-sized pores. Their chemical composition, pore dimensions, and high surface area can be tailored to yield diverse functionalities such as selective adsorption and separation, $CO_2$ capture, host-guest assembly, catalysis, mass transport facilitation, ion exchange, and acting as softeners and gas sensors, which are important in various scientific and industrial applications.[1-3]

Around 40 zeolitic frameworks are found in nature. Millions of different zeolite frameworks have been identified computationally, but only 256 were synthesized in the laboratory, as recognized by the International Zeolite Association (IZA).[4] Although new zeolitic frameworks are synthesized regularly, repeatability, stability, and optimization have been reported as major challenges because the final framework structure and functionalities are sensitive to minor variations in the synthesis parameters.[4-7] We argue that these problems are related to the energy delivery mechanism and how it drives and controls nucleation and growth.

Heat diffuses gradually into the precursor solution in hydrothermal synthesis processes. Nonuniform temperatures and reactant concentrations in liquid volume spatially alter the nucleation and growth conditions as more or less favorable. This hinders uniform particle formation and produces a relatively broad crystal size distribution.[8,9] Conversely, in microwave synthesis, energy is delivered directly to the reactants, resulting in localized heat points throughout the liquid volume. However, lacking an efficient mechanism to dissipate the excess heat, the process creates an unstable, locally superheated liquid, highly sensitive to even minor perturbations, forming low-quality zeolite crystals.[9,10] Alternatives to these conventional synthesis techniques, such as gamma-ray,[11] high-energy electron beam irradiation,[12] and ultraviolet radiation,[13] use high-energy radiation to induce ionization to form hydroxyl free radicals, lowering activation energies and accelerating nucleation and growth. However, complex processes, safety concerns, repeatability issues, and other limitations have hindered their widespread adoption.[11,12]

The ultrafast laser synthesis approach introduces an entirely different energy delivery mechanism, rendering absolute temperature and pressure values less critical. A tiny ultrafast reactor at the focus of a femtosecond laser beam can process a liquid volume of a few thousand $\mu m^3$ with extremely high spatiotemporal energy inputs and thermal gradients building up to $10^6$ K $mm^{-1}$ within several 100 fs, aligning with the timescales of silica polymerization and depolymerization reactions. Nonlinear light absorption (multiphoton) by the precursor solution generates a cascade of events that occur in rapid succession.[14,15] These include plasma generation, cavitation bubble formation, the emergence of high-speed convective flows[16,17] and shock (pressure) waves,[18,19] which facilitate controlled nucleation and growth of nearly uniform zeolite crystals. Using this technique, various zeolitic frameworks are synthesized, including MFI, LTA and FAU, with different pore sizes and configurations and crystallinities surpassing 90% with a wt.% yield of around 70%.

Uniquely, the synthesis can be stopped within fractions of a second by turning off the laser. The laser is electronically controlled and can be turned on and off within microseconds. This can be repeated as much as needed, allowing for iterative sample collection. Turning off the laser rapidly terminates all the laser-induced dynamic phenomena, halting the nucleation and growth. Since no significant residual heat remains in the system to assist further chemical reactions, the structures formed up to that point are effectively "frozen" in their self-assembled state. Samples can easily be collected for diagnostic analysis, after which synthesis can be resumed by turning on the laser. Using this capacity, the sequential evolution of the synthesis pathway of TPA-silicate-1 zeolites is followed from the formation of early oligomers to the emergence of fully grown crystals.

## 2. Results and Discussion

### 2.1. The energy delivery mechanism

The tiny ultrafast reactor is created by focusing a femtosecond laser beam at the glass-liquid interface (**Figure 1A**). This approach mirrors our previous work on quasi-2D colloidal self-assembly[16,17] where laser pulses are nonlinearly absorbed through multiphoton absorption, generating a cascade of events that drive the self-assembly of a variety of active and passive colloidal particles and living organisms.

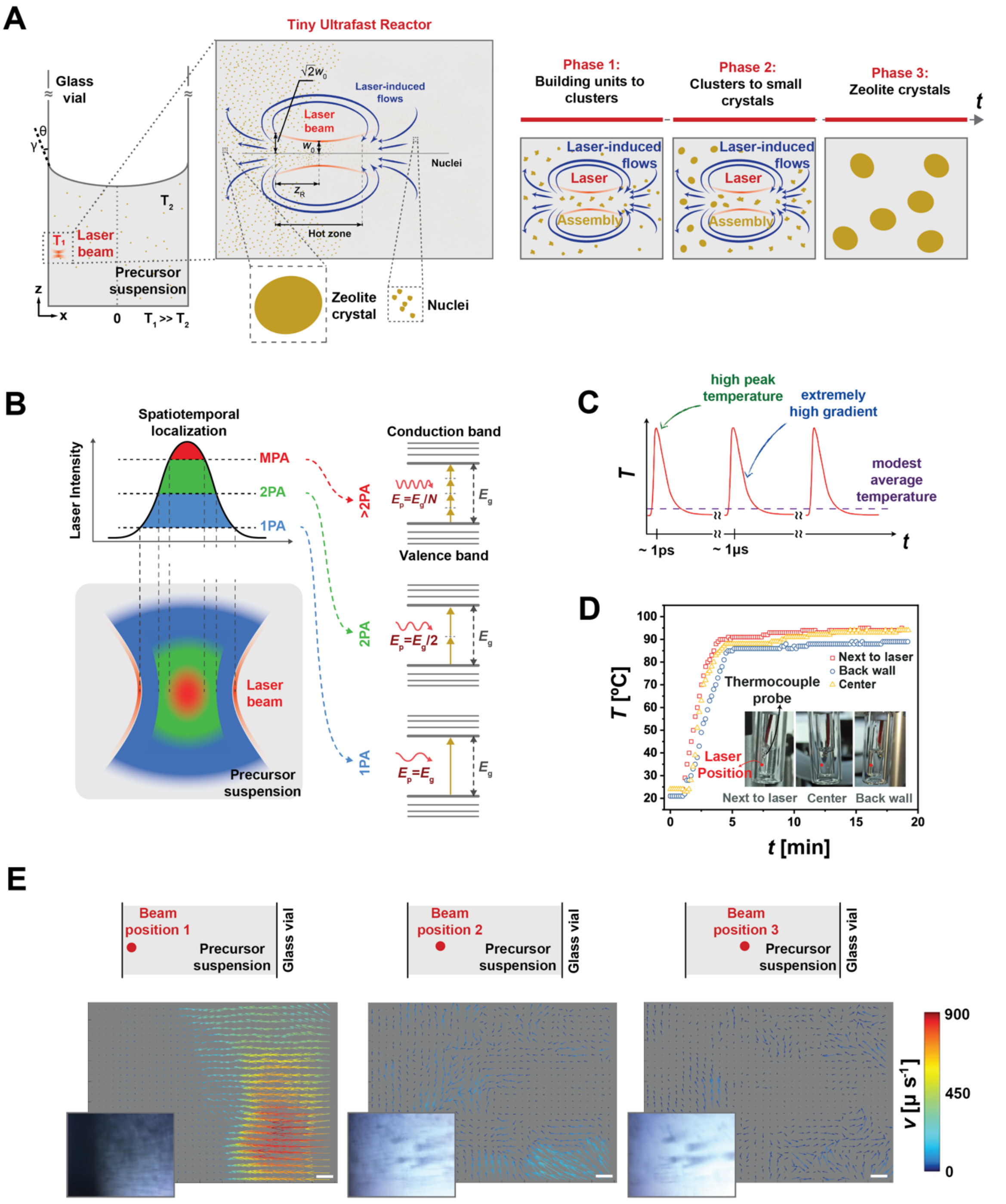

**Figure 1. (A)** Schematics illustrating the working mechanism of a tiny ultrafast reactor at the glass-liquid interface. Reactions are initiated in the hot zone near the beam waist ($\omega_0$), which is twice the Rayleigh length ($z_R$), the distance along the beam's propagation direction from the waist to the position where the cross-section area is doubled. The zeolitic crystals are produced in this tiny reactor and distributed to the full liquid volume via laser-induced convective flows, which emerge due to extreme thermal gradients and surface tensions ($\gamma$) **(B)** A simplified sketch depicts linear and nonlinear absorption processes. 1PA, 2PA, and MPA indicate the absorption of one, two, and multiple photons. 1PA (linear absorption) creates an hourglass-shaped absorption profile (blue) in the precursor solution, while MPA (nonlinear absorption) is extremely localized (red). **(C)** A cartoon-like illustration of temperature profile induced by an ultrafast pulse train. Each pulse delivers identical energy at precise intervals. Upon each pulse's arrival, the local temperature spikes almost instantaneously (within femtoseconds), creating extreme temporal thermal gradients. This temperature increase lasts only as long as the pulse, after which it begins to cool rapidly until the next pulse arrives a few microseconds later. **(D)** Recorded temperature readings with the thermocouple positioned at the front wall near the beam (position 1), the center of the bulk liquid (position 2), and the back wall away from the beam (position 3) showing that time-averaged local temperature remains steady around 95 °C. **(E)** Average laser-induced flow fields were calculated using particle image velocimetry (PIV) analysis of video recordings, where the laser beam was focused at three positions: at the glass-liquid interface (beam position 1), near the interface (beam position 2), and within the bulk liquid (beam position 3). The insets show frames captured from the video recordings. The velocity color bar shows the lower bound for the fluid velocity calculations.

Multiphoton absorption occurs when multiple photons are absorbed simultaneously by the medium, enabling an electron to transition to a higher energy state, even if the energy of each individual photon is insufficient to bridge the gap between energy levels (**Figure 1B**). This process requires high-intensity light, such as our tightly focused ultrafast laser beam, where the extremely high photon density at the focal point allows multiple photons to interact with a single atom or molecule on the timescale of its electronic response. This confines energy deposition to a significantly smaller volume (~$10^3$ μ$m^3$) than the focal volume (>$10^3$ μ$m^3$). As a result, the spatial temperature gradients exceed $10^6$ K $mm^{-1}$ within the tiny ultrafast reactor but only for a fleeting moment (in the order of the pulse duration) (see Supplementary Information). This extreme localization creates a steep spatiotemporal thermal gradient (**Figure 1C**). As a result, the time-averaged local temperature remains steady around a much lower value (**Figure 1D**).

Spatiotemporal temperature gradients, together with surface tensions at the glass-liquid interface (Figure 1A), drive and indefinitely sustain high-speed Marangoni-type convective flows (**Figure 1E**) along the beam propagation axis as long as the laser beam is incident. These flows can be clearly observed in real-time experimental recordings (**Video 1** and **2**). A simple calculation (see Supplementary Information) shows that the entire chemical reaction occurs on a microsecond timescale. We assume that the local temperature remains substantially above the average temperature of 95 °C only within the interaction volume and that the precursor solution is mixed well enough during the experiment. The total number of pulses during a typical experiment of 240 min. and at 200 kHz repetition rate is 3 billion. The ratio of the interaction volume to the total solution is similarly about 0.4 billion. Hence, any given portion of the precursor solution is processed at high temperatures for the timespan of about 10 pulses, or about 50 μs.

The working dynamics of this unique energy delivery mechanism in self-assembling zeolitic crystals are discussed in the next section. However, before proceeding, we would like to emphasize the unique role of ultrafast lasers in energy delivery.

Historically, significant progress has been achieved in the ultrafast synthesis of colloidal nanoparticles in liquid,[20] where the nanoparticles are ablated from a solid target immersed in the liquid. Subsequent hydrothermal synthesis is typically used following the laser treatment to grow the nuclei into mature crystals.[21] The potential of ultrafast lasers in synthesizing zeolites (or other inorganic nanomaterials) by driving chemical reactions directly and fully in liquid has not been thoroughly explored. We believe that the attempts to process the entire precursor solution in a single step likely faced two major challenges: (1) Insufficient energy deposition to the solution, a challenge compounded by the lack of detailed parameters in existing literature, such as pulse fluence and repetition rate (by contrast, see our optimization analysis in **Figure 2**, supplemented by **Table S1**). (2) The difficulty in initiating nonlinear light-matter interactions necessary for processing the entire liquid volume. This issue, while not always explicitly stated, is suggested by experimental setup diagrams, where the beam is often pointed (which might not be tightly focused) arbitrarily within the bulk liquid, which is suboptimal for efficient multiphoton absorption by the solution.

We conducted further experiments to support arguments on the second challenge. A typical synthesis process captured in Videos 1 and 2 visualizes the flow fields at three distinct beam positions: at the glass-liquid interface (left), in close proximity to the interface (middle), and closer to the centre of bulk liquid (right). An average speed is computed over a smaller time frame to highlight the distinct characteristics of the flow fields for the three beam positions, which appear at the end of Video 2 and are shown in Figure 1E. Significantly, high-speed flows manifest exclusively when the beam is focused at the glass-liquid interface, whereas seemingly unstructured slow flows are observed away from the interface. This is expected as surface tensions (Figure 1A) promote the creation of Marangoni flows at the interface,[16,17] a condition absent at the two beam positions away from the glass walls. Moreover, the transition of the precursor suspension color from transparent to opaque white in Video 1 signifies the saturation of the liquid volume with fully-grown crystals, which is exclusively achieved when the beam is focused at the interface under the same reaction time.

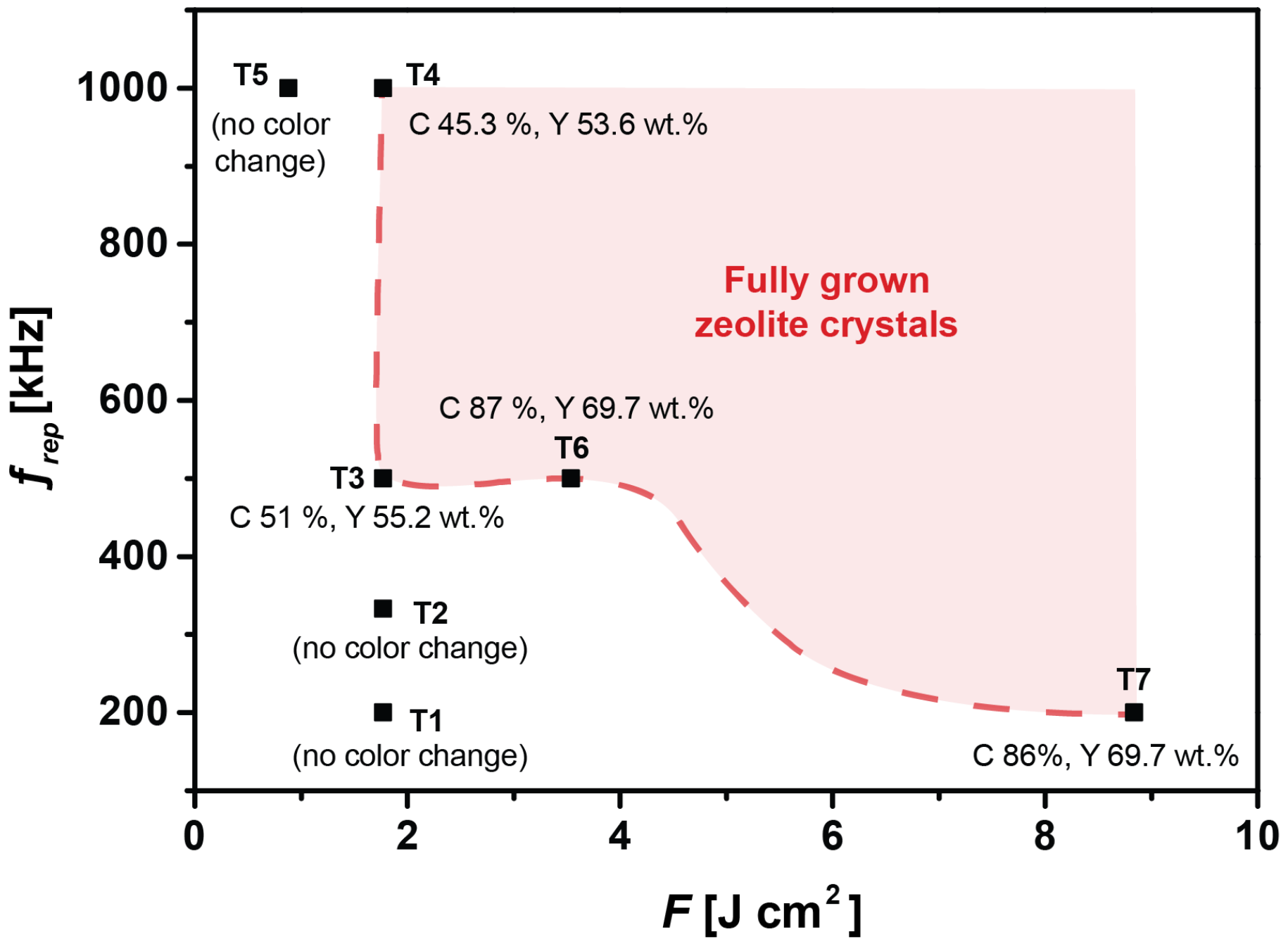

**Figure 2.** Graph depicting the optimal laser parameters for the formation of fully-grown zeolite crystals. The repetition rate ($f_{rep}$) increases while maintaining the total deposited energy and pulse fluence ($F$) constant (T1, T2, T3, T4). $F$ is halved while keeping the total deposited energy and $f_{rep}$ unchanged (T5). $F$ is doubled while maintaining the total deposited energy and $f_{rep}$ constant (T6). $F$ is significantly increased while keeping $f_{rep}$ constant (T7). The crystallinity (C) and weight percentage yield (Y) values are provided for comparison in each trial.

Mechanical stirring of the precursor solution in conventional hydrothermal synthesis has been studied by gradually rotating the autoclave, typically within a speed range of 10-20 rpm.[22,23] This approach has yielded positive outcomes, such as slightly accelerated crystal growth and a relatively narrow size distribution. We further asked if the benefits reported here could be solely attributed to the laser-induced convective flows. To examine this, the laser beam is deliberately directed onto the bulk liquid (avoiding the glass-liquid interface) to prevent the formation of Marangoni flows along with all the dynamic phenomena resulting from multiphoton absorption of pulses. Instead, the solution is mechanically mixed using a magnetic stirrer (**Video 3**). As anticipated, no noticeable change in color was observed within the precursor solution, even after tripling the laser exposure time compared to the typical laser synthesis conditions reported here.

As for using other lasers for syntheses, such as continuous-wave (CW) and nanosecond (ns), we argue that the attempts would likely fail, as these lasers cannot replicate the dynamic conditions (e.g., flows, bubbles, etc.) that femtosecond lasers achieve. CW lasers function more like inefficient furnaces, gradually heating the precursor solution, which only slowly promotes nucleation and growth. Although ns lasers can induce multiphoton absorption, they require significantly higher pulse fluences due to their longer pulse durations and lower peak intensities. This extended pulse duration causes thermal diffusion, leading to heat buildup in the material, which can trigger undesirable thermal effects, even melting or ablating the glass vial. This compromises precision, especially in applications requiring spatiotemporal control.

**2.2. The self-assembly mechanism**

The initial phase involves initiating silica polymerization-depolymerization reactions within the hot zone of the laser beam (Figure 1A), effectively serving as a tiny ultrafast reactor for zeolite synthesis. The local temperature in the reactor is periodically elevated, momentarily reaching extremely high levels that significantly accelerate the chemical reaction.

Early-stage oligomer rings and condensed units, produced with each laser pulse, are systematically dispersed throughout the precursor solution via convective flows. The reactor volume is on the order of several thousand $\mu m^3$, less than a billionth of the total liquid volume. As a result, any self-assembled unit reaches thermal equilibrium with the bulk long before it re-enters the reactor. This exceptionally small reactor-to-bulk volume ratio ensures that, while local temperatures within the reactor are extremely high, the average temperature across the entire system remains low, around the measured value of 95 °C (Figure 1D).

The energy input by the femtosecond laser pulses is not only used in chemical synthesis but also in generating and sustaining laser-induced dynamic phenomena. A rapid buildup of free electrons upon multiphoton absorption triggers optical breakdown. This results in the formation of a transient plasma at extremely high temperatures and pressures. Plasma increases the concentration of reactive species and further promotes nucleation and growth. As the plasma expands, it displaces the surrounding liquid, generating cavitation bubbles. Together with laser-induced convective flows, these buoyant bubbles enhance the efficient mixing of the precursor solution, promoting faster and nearly uniform crystal growth. The bubbles grow due to the extreme pressure difference. Some subsequently collapse, emitting shock waves that further influence the surrounding medium. Extreme pressures propagate, further assisting nucleation and growth. This sequence of events plays a critical role in energy transfer, fluid dynamics, and chemical reactions within the tiny ultrafast reactor.

As the reactor is the only hot zone and the bulk liquid is significantly colder, convective flows continually move between the reactor and the bulk liquid. This is crucial in continually transporting previously distributed nuclei and particles back to the reactor, where they grow further during each iteration, evolving into larger structures and maturing into nearly mono-sized and discrete zeolite crystals (see Supplementary Information). This unparalleled efficiency sets ultrafast laser synthesis apart from other methods in the field.

Another great benefit of ultrafast laser synthesis in controlling self-assembly stems from the dynamic control over the rate of energy delivery — a capability absent in both conventional (such as hydrothermal and microwave methods) and recently developed approaches (including assembly-disassembly-organization-reassembly,[24] seed-directed secondary growth,[25] ultraviolet,[13] gamma-ray,[11] and high-energy electron beam irradiation[12]. This is achievable by adjusting pertinent laser parameters such as average laser power, peak power, and pulse energy fluence.

Furthermore, the process can be halted and resumed by turning the laser on and off (**Video 4**) for rapid cooling and heating of the liquid medium, exerting a profound influence on the kinetics of the chemical reaction. Rapid cooling is possible because laser energy is delivered only to the focal volume, and excess heat is immediately carried out via convective flows and distributed to the bulk liquid. If the laser beam is turned off, no significant heat source remains to support further nucleation and growth (see Supplementary Information). That way, the self-assembled structures up until that point are effectively frozen and preserved. Reintroducing the laser pulses rapidly provides the energy to promote the chemical reactions and resume the self-assembly process.

### 2.3. Characteristics of ultrafast laser synthesized zeolites

A selection of laser-synthesized zeolites is shown in **Figure 3**, with their crystalline identities confirmed via XRD spectroscopy (see also **Figure S3** and **S4** for MFI-type ZSM-5 and for FAU-type Zeolite-Y). To ensure the reliability and reproducibility of the synthesized zeolites, we conducted over 120 experiments, consistently yielding similar results (see **Figure S5**, **S6**, and **S7** for the results of representative experiments).

Our focus centered on TPA-silicalite-1, a well-documented and extensively studied model zeolite. To draw quantitative comparisons, half of the prepared precursor solution was subjected to laser synthesis for 3 hours, while the other half underwent a typical hydrothermal synthesis for 48 hours. Comparative analysis of the average crystal size distribution between laser- and hydrothermal-synthesized zeolites reveals distinct outcomes.

Laser synthesis yields a narrow distribution with larger crystal sizes (**Figure S8A** and **S8C**). In contrast, hydrothermal synthesis results in a broader distribution featuring relatively smaller ones (**Figure S8B** and **S8D**). Notably, akin to hydrothermal synthesis, reducing the average crystal size in laser synthesis is feasible by modifying the molar formula's water content from M1 (Figure S8) to M2 (**Figure S11**) and M3 (**Figure S10**) (see also **Figure S11** and **Table S2**).

The relatively large crystal sizes and narrow size distribution of ultrafast laser-synthesized zeolites result from the unique energy delivery mechanism discussed in the previous two sections. Heat and the reactants diffuse and transport significantly more slowly in conventional hydrothermal synthesis, creating uneven nucleation and growth conditions resulting in larger size variations.

Thermal stabilities, as measured through thermogravimetric (TGA) and differential thermal (DTA) analyses for both laser and hydrothermally synthesized (**Figure S12**) zeolites, yield comparable outcomes with water removal of 4.2% at around 109 °C in the case of laser and 3.4% at around 76.4 °C in hydrothermal synthesis and removal of 12.1% TPA content for both methods at around 334 °C in the case of laser and 329 °C in hydrothermal synthesis.

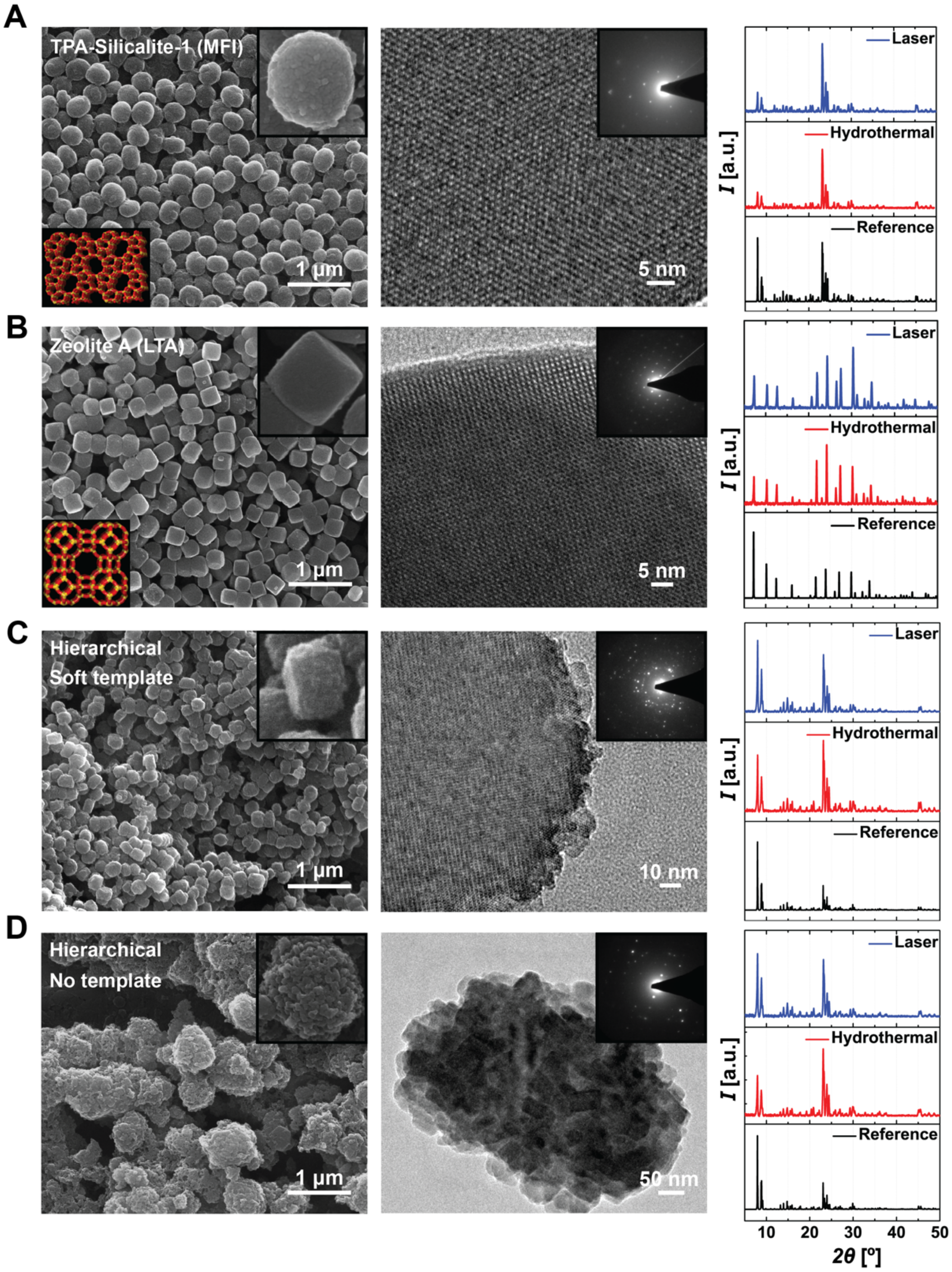

**Figure 3.** SEM (left) and TEM (middle) images of **(A)** TPA-Silicalite-1 (MFI type), **(B)** Zeolite A (LTA type), **(C)** hierarchical ZSM-5 with a soft template, and **(D)** mesoporogen-free hierarchical ZSM-5 zeolites. The left panel insets show unit cell structures (lower left) and SEM images of individual crystals (upper right). The middle panel insets show SAD patterns. XRD spectrums of the laser- (blue, upper) and hydrothermal-synthesized (red, middle) zeolites are compared to their International Zeolite Association (IZA) references (black, lower layer) at the rightmost panel.

Crystallinity and product yield calculations suggest similar results for laser- and hydrothermally synthesized zeolites (Table S2). $N_2$ adsorption-desorption curves (**Figure S13**) manifest a Type I isotherm, indicative of a microporous structure (pore size <2 nm)[29] with the Brunauer-Emmett-Teller (BET) surface area of 335.5 $m^2$ $g^{-1}$ (falling within the typical range of 332 - 400 $m^2$ $g^{-1}$)[30-33] for laser synthesized TPA-silicalite-1 crystals. In accordance with IZA reported values, DFT analyses show that the pore volume and effective pore radius of the laser-synthesized TPA-silicalite-1 crystals are 0.164 $cm^3$ $g^{-1}$ and 2.88 Å, respectively (**Figure S14**).

### 2.4. Nucleation and growth pathways

We sampled the process at intervals from the 20th to the 300th minute by turning the laser off and on and identified two distinct regimes for nucleation and growth. Throughout the synthesis, no observable color change occurred in the precursor solution until the 70th minute. The initially light milky color at 70th min. transformed to a milky hue by the 90th min., later evolving into an opaque white appearance by the 240th min. (**Figure 4A**).

Subsequent SEM images (**Figure 4B**) and average crystal size distributions (**Figure 4C**) reveal where a high-degree network of smaller crystals with rough surfaces gradually transitioned into larger, discrete crystals with smoother surfaces. SEM and TEM images of samples collected at the 70th min. suggest that individual crystals are formed by aggregating smaller crystals, each with an average size of ~15 nm. Since those on the surface and contributing to the network by bridging crystal aggregates are electron-beam sensitive, phase-contrast TEM images were performed, revealing a bright appearance and confirming their crystalline nature (**Figure 4D**).

Moreover, to elucidate the chemical transformations occurring during the formation of these minute crystals, we scrutinized the Si-O-Si and O-Si-O bending and the Si-O-Si symmetric stretching signals using ATR-FTIR spectroscopy. Comparative analysis with molecular dynamics simulations[31-33] (**Figures 5A** and **5B**) reveals the presence of loosely connected 5-membered ring structures (yellow-filled peak), condensed units of 10T and 22T structures (red-filled peak), and 36T MFI precursors (brown-filled peak) as early as 20th minute of the synthesis. Fully grown MFI structures appear at the 60th minute (grey-filled peak) (see also **Figure S15**).

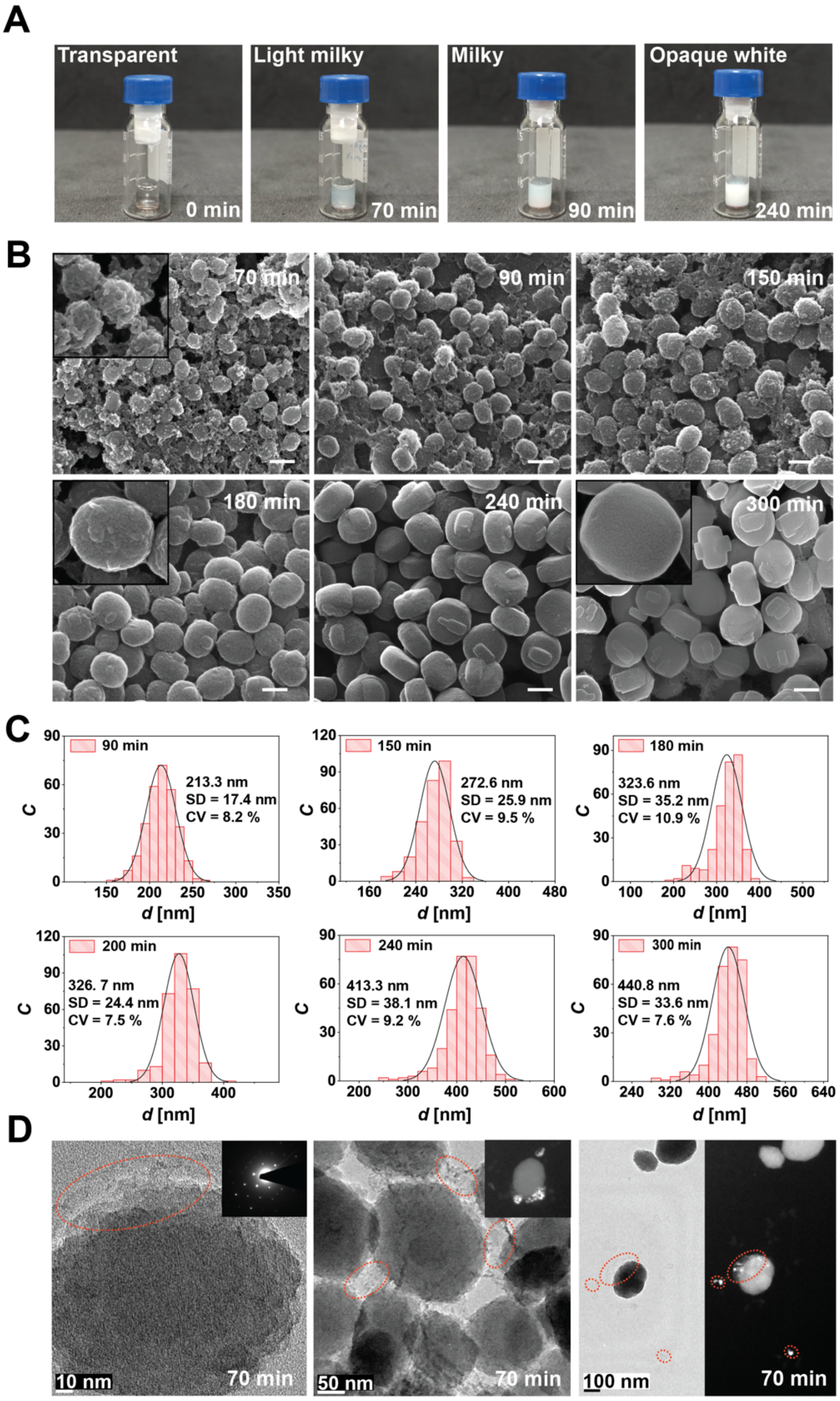

**Figure 4. (A)** Photographs taken at different stages of laser synthesis show color change from transparent to milky white and opaque white as the reaction time increases. **(B)** SEM images show the evolution of crystals from rough to smoother surfaces as the reaction time increases. **(C)** Average crystal size distribution plots for zeolites synthesized in various reaction times. *C* denotes particle counts. **(D)** TEM images of crystals were obtained after $70^{th}$ min. reaction

time shows rough surface contours (dashed red ellipses). The phase contrast TEM images show that these marked regions appear bright, confirming their crystalline nature.

The evolution of peak relative intensities is tracked until the 300th minute of the reaction. Initially, the signal from early ring vibrations diminishes until the 70th minute and then ceases at the 90th minute, marking the complete consumption of small oligomers. Concurrently, the signal contribution from MFI precursors increases as they assimilate the early oligomers for their self-assembly. After the 70th minute, the signal from fully grown MFI structures increases drastically by consuming the MFI precursors. Accordingly, the signal contribution of the MFI precursors diminished significantly.

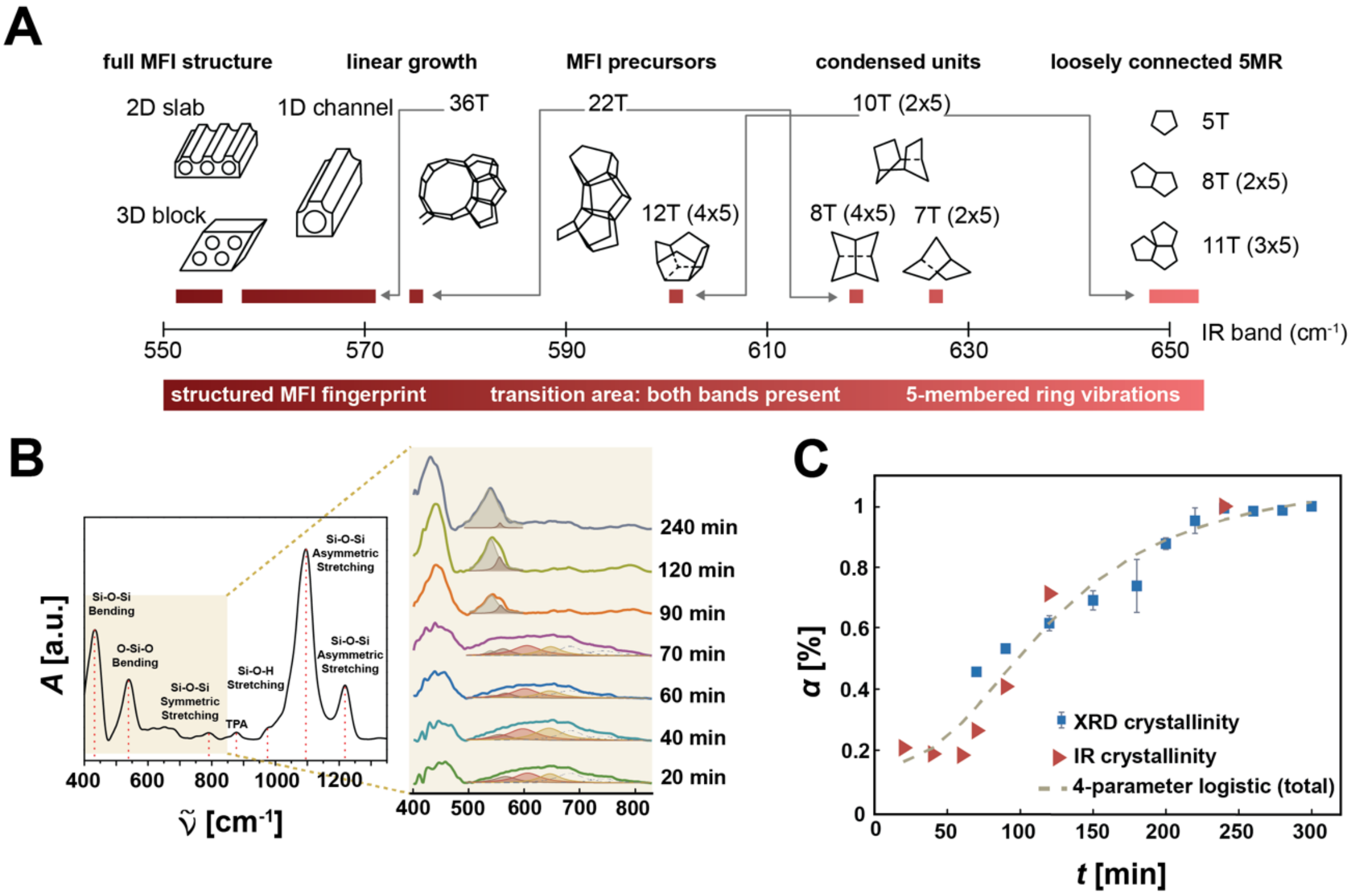

**Figure 5. (A)** Molecular dynamics simulations show the IR band shift corresponding to different nucleation and growth stages of TPA-Silicalite-1 zeolites. Reproduced with permission.[31] 2008, American Chemical Society. **(B)** The IR absorbance spectrum of zeolites with signal contributions from loosely connected 5-membered ring structures (yellow-filled peak centered at wavenumber ~650 cm$^{-1}$) to condensed units of 10T and 22T structures (red-filled peak centered at ~600 cm$^{-1}$), 36T MFI precursors (brown-filled peak centered at ~560 cm$^{-1}$), and fully grown MFI structures (grey-filled peak centered at ~540 cm$^{-1}$). **(C)** The graph shows the sigmoidal growth of the crystallinity index ($\alpha$) calculated from ATR-FTIR (red triangles) and XRD (blue squares) to synthesis time.

We calculated the crystallinity of liquid samples collected at $20 < t < 300$ minutes of the synthesis from XRD data and also from ATR-FTIR analyses (**Figure 5C**) to correlate them with the self-assembled structures calculated through the molecular dynamics simulations (Figure 5B). Our crystallinity analyses revealed fast and slow sigmoidal growth dynamics for MFI precursor structures and fully-grown crystals, respectively.

The consumption of early oligomers by the growing MFI precursor structures until the $90^{th}$ min. was much faster than that of MFI precursors by fully-grown zeolite crystals. This is so because strong nonlinear light-matter interactions that drive the nucleation and growth are modified when the precursor solution transitions from transparent to opaque white, affecting the growth rate. Multiphoton absorption is the dominant absorption mechanism when the medium is transparent, and we start our experiments with a transparent precursor solution rather than its gel-like form, typically used in conventional hydrothermal synthesis.

The initial transparent solution turns milky white at around 90 min. (Figure 4A), corresponding to where the initial fast growth rate slows in the crystallinity graph in Figure 5C. The slower growth rate eventually plateaus around 240 min. when the precursor solution is opaque white, indicating strong scattering.

**2.5. Versatility of the ultrafast laser synthesis technique**

Since femtosecond laser pulses provide energy within the relevant timescales of zeolite polymerization-depolymerization reactions, it becomes feasible to synthesize zeolites without the customary aging step. Typically, in zeolite synthesis, the precursor solution undergoes a standard practice of being kept at room temperature for 24 hours post-chemical mixing. To show that this step might not be necessary, we applied the laser immediately after chemical mixing and successfully synthesized TPA-Silicalite-1 crystals (**Figure 6A** and **6B**).

Another distinctive feature of utilizing ultrafast lasers is the potential to leverage collective pulse-matter interactions for synthesizing diverse topologies and functionalities, thereby further exploring the configurational phase space of zeolites. This is a new interaction regime with different kinetics whereby multiple pulses simultaneously interact with the matter instead of a single pulse. To experiment with this idea, we utilized a home-built ultrafast burst-mode laser.[34] **Figure 6C** shows that our initial findings with this new light-matter interaction regime produced hierarchical structures with pore openings spanning approximately 20 to 60 nm and 120 to 280 nm.

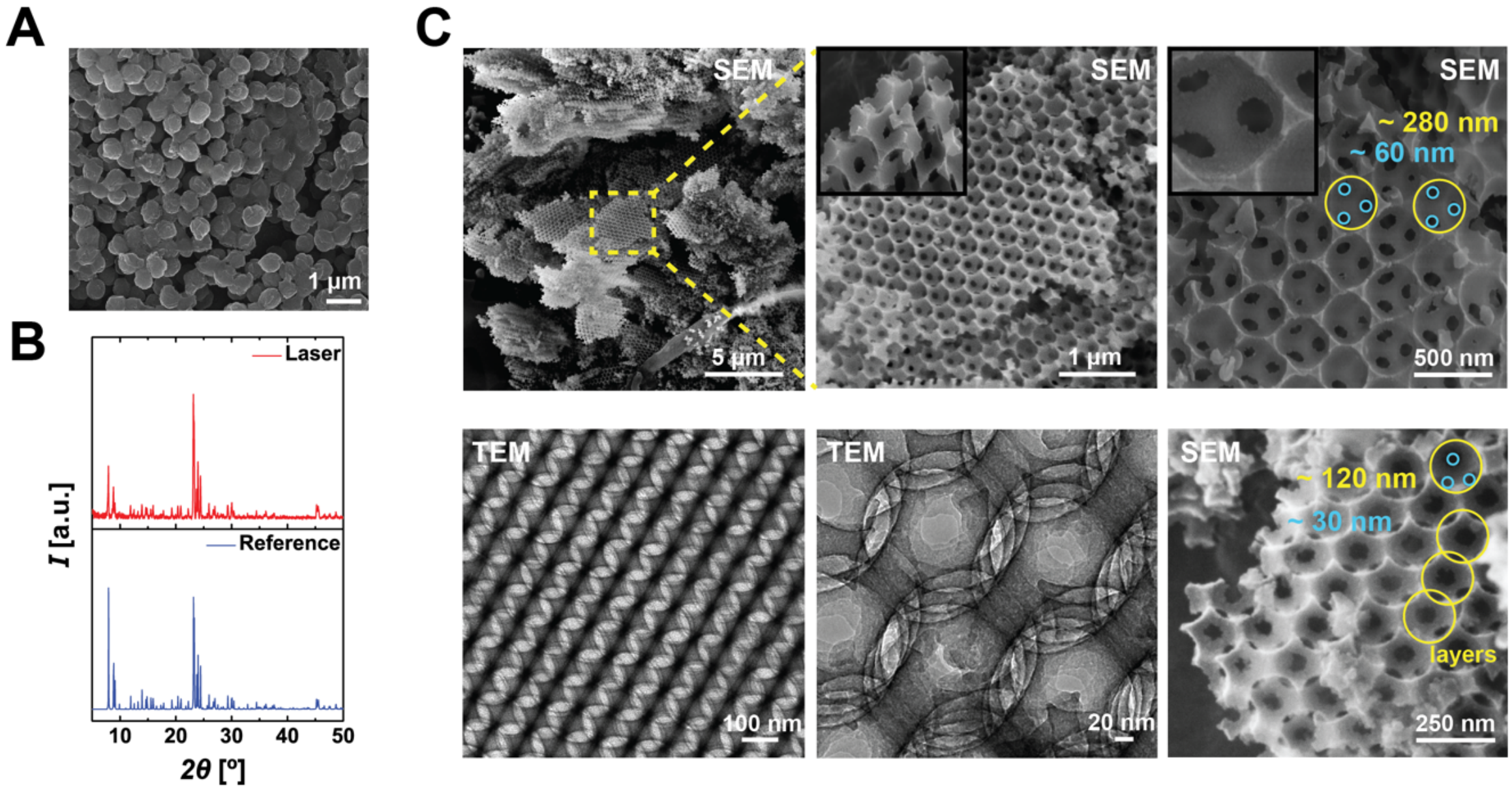

**Figure 6. (A)** SEM image and **(B)** XRD pattern showing laser-synthesized TPA-Silicalite-1 without the preliminary aging step. **(C)** SEM and TEM images showcase laser-synthesized hierarchical structures.

## 3. Conclusions

The ultrafast laser synthesis technique reported here is a potential paradigm shift in zeolite synthesis. Femtosecond laser pulses provide extremely localized energy in space and time, allowing high-precision control over energy delivery, reaction kinetics, and chemical synthesis. Energy is delivered at timescales compatible with the natural timescales of zeolite polymerization and depolymerization reactions, which makes it possible to even eliminate the time-consuming aging step. High spatiotemporal precision prevents unwanted heat accumulation and thermal diffusion, promoting ultrafast synthesis of nearly uniform, high-quality, high-yield crystals. The technique opens new possibilities for producing diverse zeolitic structures and topologies. Various zeolite configurations may be explored by adjusting laser parameters without altering the chemical composition. Additionally, collective light-matter interactions such as the one facilitated by burst-mode lasers can be employed to explore more complex topologies that may be difficult or impossible to achieve using conventional methods, such as millions of computer-generated zeolitic frameworks. We postulate that the technique introduced here is not limited to zeolite synthesis as laser-induced dynamic phenomena driving the chemical synthesis are not specific to the chemicals used in the process. Synthesis of a broad range of inorganic materials should be possible.

## 4. Methods

*Experimental Setup*: The precursor solution is housed in a glass insert (Supelco, Merck, 0.75 mL), which is encased within a chromatography glass vial (IsoLab, 4 mL, N13) securely positioned using an adjustable sample holder attached to its cap (Video 1). An ultrafast laser beam (Spectra-Physics, Spirit One, 1040 - 8 - SHG) is focused at the glass-liquid interface. The laser operates at a central wavelength of 1040 nm, with a pulse duration of 300 fs and a repetition rate of 200 kHz. To generate Figure 2, we either fixed the total deposited energy or the pulse fluence to investigate the impact of laser parameters on zeolite formation. Consequently, we varied the average laser power from 2.5 W to 5 W and adjusted the repetition rate between 200 kHz and 1 MHz for these experiments.

*Video Recordings and Analyses*: The experiments were captured using CMOS scientific (DCC1645C-HQ, Thorlabs) and Canon DSLR (Eos) cameras, employing a 10× Nikon objective for imaging. A white light source was positioned on the optical table at an appropriate angle to illuminate the glass vial and enhance visibility. Particle Image Velocimetry (PIV) analyses were conducted on recordings with a frame rate of ~30 fps and a resolution of 1116 × 891 pixels, utilizing DynamicStudio Software from Dantec Dynamics over ~420 frames. Grayscale images of velocity fields were generated for adaptive PIV analyses to calculate the 2D velocity fields on a 2D grid with tiles sized 16 × 16 pixels. The average displacements in the $x$- and $y$-directions were then extracted using the software's MATLAB link option and converted to velocity vectors in units of $\mu m\ s^{-1}$. These vectors were colored using MATLAB.[35] Velocity fields were then averaged over time to derive the mean field on the sample within the specified time interval. Real-time video recordings of the experiments clearly show that the flows are much faster than the calculated values. This discrepancy arises because the velocity fields are determined by detecting and tracking buoyant cavitation bubbles, which vary in size and are carried by high-speed convective flows. These bubbles also move rapidly in and out of the focal plane, making it prohibitively difficult to track them consistently. Since the bubbles do not fully capture the fluid flow dynamics, the actual fluid velocities are expected to be significantly higher than the average velocities calculated from bubble movement (see also Supplementary Information).

*Temperature Profiling of the Suspension:* The suspension temperature was measured using a K-type thermocouple probe inserted in the liquid during a typical synthesis. The temperature values were measured using a digital multimeter (Protek 506) connected to a computer to record the measurements for 20 minutes. The laser beam was blocked within the first 2 minutes of the measurement.

*List of Chemicals used in this work*:

- Aluminium isopropoxide ($Al(iPro)_3$, > 98%, 220418, Aldrich),
- Al powder (325 mesh, 99.5%, Alfa Aesar),
- Deionized water (DI $H_2O$, resistivity = 18.2 MΩ),
- Ethanol (EtOH, 99.99%, IsoLab),
- Sodium hydroxide (NaOH, 99%, Merck),
- Tetraethyl orthosilicate (TEOS, 98%, 131903, Aldrich),
- Tetramethylammonium hydroxide (TMAOH, 25 wt.%, Aldrich),
- Tetrapropylammonium hydroxide (TPAOH, 1M, 254533, Aldrich),
- Cetyltrimethylammonium bromide (CTAB, 98%, Fisher Chemical),
- $SiO_2$ (Ludox-HS 30, 30 wt.% $SiO_2$, pH=9.8, Aldrich), and
- Polystyrene spheres (Micro Particles GmbH, diameter: 250 nm & 500 nm, 2 wt.%).

*Synthesis Procedures*:

*TPA-silicalite-1 zeolite (MFI type):* Three different molar formulas[36,37] were used for MFI-type TPA-Silicalite-1 zeolite synthesis:

$$M1 = 25\ SiO_2 : 9\ TPAOH : 1450\ H_2O : 100\ EtOH$$

$$M2 = 25\ SiO_2 : 9\ TPAOH : 480\ H_2O : 100\ EtOH$$

$$M3 = 25\ SiO_2 : 9\ TPAOH : 450\ H_2O : 100\ EtOH$$

We primarily used the M1 and only used the M2 and M3 formulas to demonstrate their potential in reducing average crystal sizes by minimizing water content. To prepare the precursor suspension, TPAOH (0.25 g) was combined with TEOS (0.142 g) in a bottle and stirred at room temperature (RT) for 30 min. Subsequently, DI water (0.512 g) was added to this solution and stirred at RT for 24 h. The same synthesis procedure was applied to the M2 and M3 molar formulas, with DI water (0.07 g) utilized for M2 and no additional water for M3. Following preparation, the solution was divided into two batches for ultrafast laser and hydrothermal synthesis to compare the resulting zeolite crystals. Hydrothermal syntheses were conducted according to recommended procedures. Hydrothermal synthesis for batches that use M1 molar formula were conducted at 100 °C in an oven (Binder FD 115) for 48 h. These conditions changed to 90 °C heating for 30 h according to synthesis requirements using M2 and M3 molar formulas. Powder products were collected following centrifugation (14000 rpm, Eppendorf Centrifuge 5420), washed with DI water until reaching a neutral pH, and dried overnight at 45 °C in the oven.

*Zeolite A (LTA type):* $Al(iPro)_3$ : 3 TEOS : 7.36 TMAOH : 0.33 NaOH : 192.41 $H_2O$ molar formula was used to synthesize microporous zeolite A. TMAOH (2.17 g) was combined with DI water (0.95 g) and stirred for 20 min. Subsequently, $Al(iPro)_3$ (0.17 g) was gradually added to the mixture and stirred for 1 h. After this, TEOS (0.53 g) was introduced to the mixture and stirred for 2 h. Meanwhile, NaOH (0.01 g) and DI water (0.23 g) were mixed in a separate beaker and slowly added dropwise to the main mixture. The resulting blend was stirred for 15 h at RT for aging. The laser synthesis lasted 4 h, while the hydrothermal reaction was conducted in an oven at 100 °C for 8 h. Powder products were collected using the same procedure as that of TPA-Silicalite-1 zeolites.

*ZSM-5 zeolite (MFI type):* $Al(iPro)_3$ : 50 TEOS : 9 TPAOH : 9 NaOH : 5709 $H_2O$ molar formula[38] was used to synthesize microporous ZSM-5 zeolites. DI water (0.42 g) was combined with TEOS (0.26 g), followed by the slow addition of $Al(iPro)_3$ (0.005 g), forming Solution A. Meanwhile, in a 5 mL beaker, NaOH powder (0.009 g) was dissolved in DI water (1.92 g) and stirred for 2 min. until fully dissolved. This NaOH solution (Solution B) was mixed with TPAOH (0.23 g). Solution A and Solution B were stirred at RT for 1 h. Subsequently, Solution B was added dropwise to Solution A, stirring the mixture for an additional 24 h at RT for aging. The laser synthesis lasted 6 h, while the hydrothermal reaction was conducted in an oven at 100 °C for 24 h. Powder products were collected using the same procedure as that of TPA-Silicalite-1 zeolites.

*Zeolite Y (FAU type):* 9 $Na_2O$ : 0.7 $Al_2O_3$ : 10 $SiO_2$ : 160 $H_2O$ molar formula[6] was used to synthesize template-free zeolite Y. Solution A was prepared by dissolving NaOH (2 g) in DI water (4 g), to which aluminum powder (0.189 g) was gradually added. In Solution B, colloidal silica (10 g) was combined with NaOH (1.6 g) and DI water (3.4 g) to form a turbid suspension. This mixture was then subjected to an oven at 100 °C for 6 min. to convert the turbid suspension into a clear one. Afterwards, Solution A was added drop by drop into Solution B under vigorous stirring, with Solution B cooling down during the mixing process.

The resulting clear suspension was left to age 24 h at RT. The laser synthesis procedure lasted 5 h, while the hydrothermal reaction occurred in an oven set at 100 °C for 45 h. The powder products were collected using the same procedure as that of TPA-Silicalite-1 zeolites.

*Hierarchical ZSM-5 zeolite with a soft template:* $Al(iPro)_3$ : 50 TEOS : 18 TPAOH : 1.37 CTAB : 171.5 EtOH : 4903 $HO_2$ molar formula[38] was used to synthesize hierarchical ZSM-5 zeolites using the soft template CTAB. The primary precursor suspension was aged in an oil bath at 100 °C for 44 h. CTAB, ethanol, and DI water were introduced to the primary precursor suspension. In a separate bottle, a mixture of TEOS (1.56 g) and DI water (2.59 g) was stirred at RT for 2 h, with $Al(iPro)_3$ (0.03 g) added gradually. Then, TPAOH (2.76 g) was incorporated into the solution, forming Solution A, which was subjected to an oil bath at 40 °C for 2 h. The temperature was subsequently raised to 100 °C to continue stirring the first precursor suspension for an additional 44 h. Meanwhile, CTAB (0.075 g) was dissolved in DI water (8.45 g) in a different bottle and stirred at RT for 15 min. to form Solution B. After 44 h of stirring, Solution B was added to Solution A after cooling to RT. The mixture was stirred in an oil bath at 80 °C for another 2 h. Finally, ethanol (1.19 g) was introduced to the final precursor suspension for further stirring at 80 °C for 2 h. The laser synthesis lasted 5 h, while the hydrothermal reaction occurred in an oven set at 150 °C for 24 h. The powder products were collected with the additional calcination step at 490 °C for 5 h using Protherm equipment with a heating rate of 5 °C $min^{-1}$.

*Mesoporogen-free hierarchical ZSM-5 zeolite:* $Al(iPro)_3$ : 50 TEOS : 9 TPAOH : 9 NaOH : 5709 $H_2O$ molar formula[38] was used to obtain hierarchical zeolite without using a template to obtain mesopores (*i.e.*, mesoporogens). Solution A was prepared by mixing DI water (0.83 g) with TEOS (0.51 g), then by gradually adding $Al(iPro)_3$ (0.01 g). The resulting mixture was stirred at RT for 1 h. Meanwhile, Solution B was prepared concurrently, consisting of NaOH powder (0.018 g) dissolved in DI water (0.96 g) for 2 min. until complete dissolution, followed by adding TPAOH (0.45 g). Solution B was stirred at RT for 10 min. Subsequently, Solution A was placed in an oil bath at 40 °C, and Solution B was added dropwise while stirring for 3 h at the same temperature. The oil bath temperature was then increased to 100 °C, and the solution was stirred for an additional 44 h. Afterwards, the suspension was removed from the oil bath, and DI water (2.88 g) was added to achieve the final precursor suspension. This final solution was stirred at RT for 15 min. The laser synthesis lasted 5 h, while the hydrothermal reaction occurred in an oven set at 150 °C for 24 h. The powder products were collected with the additional calcination step at 490 °C for 5 h using Protherm equipment with a heating rate of 5 °C $min^{-1}$.

*Multiporous hierarchical crystals with a hard template:* M2 molar formula of MFI-type TPA-Silicalite-1 zeolites were used to synthesize hierarchical crystals. The precursor preparation and aging steps followed the procedure outlined for TPA-Silicalite-1 zeolite synthesis, with one key variation: adding 250 nm and 500 nm polystyrene (PS) nanoparticles (1.3 mg) to precursor suspension (500 µl) in separate syntheses. These PS nanoparticles, obtained in powder form, served as hard templates to induce the formation of multi-porous topologies. To achieve the PS in powder form, the PS solution underwent centrifugation (14000 rpm, Eppendorf Centrifuge 5420) followed by overnight drying at 45 °C in an oven. Our synthesis employed a home-built ultrafast burst mode laser,[37] with each burst (groups of pulses) lasting 200 ns, intra-burst repetition rate of 1.6 GHz (correspondingly, 320 pulses per burst) and inter-burst repetition rate of 200 kHz, and pulse duration of approximately 300 fs. The laser synthesis process lasted 3 h, with the average laser power incident on the glass measured at 3 W. The powder products were collected with the addition of a calcination step performed at 500 °C for 5 h using Protherm equipment with a heating rate of 5 °C $min^{-1}$.

*Characterizations:*

*The crystallographic structure* of the zeolite powder samples was analysed using an XRD spectrometer (Malvern Panalytical X'Pert Pro Multi-purpose Diffractometer), with a Cu $K_\alpha$ X-ray source ($K_\alpha$ = 1.54187 Å) operating at 45 kV and 40 mA, scanning between the 2θ angle range of 5° - 50°. HR-TEM images and SAED patterns were obtained using a FEI Tecnai G2 20 S Twin microscope operating at 200 kV. SEM micrographs were obtained using an FEI Quanta 200F microscope operating at 30 kV after coating the samples with a 10 nm Au/Pd conductive layer using a Gatan 682 Precision Etching Coating system. FEI FP2067/30, equipped with a through-lens detector in immersion mode, was used for the FIB-SEM micrographs with higher magnifications.

*Particle size distribution analyses* were conducted using ImageJ and Origin(Pro) software, based on a minimum of four SEM images of the same sample taken from various locations along the substrate. A fixed number of crystals (298) were utilized for the analyses, with the analysis-statistics toolbox of Origin employed to generate histograms depicting the average size distribution. The results were presented as mean ± standard deviation (SD), and each form of zeolite synthesis was carried out separately at least five times. The figure captions provide the sample sizes for each particle size distribution.

*The surface area and pore size distribution* were evaluated through BET $N_2$ adsorption-desorption isotherms at 77.3 K utilizing the Quantachrome Autosorb IQ2 MP gas sorption system. A total of 18 synthesis cycles were conducted to obtain the requisite 150 mg powder sample. Before BET measurements, the powder sample underwent outgassing at 300 °C for 16 h. The BET method facilitated the determination of the total surface area, while the micropore volume and the pore size distribution were calculated via DFT.

*Thermal stability*, zeolitic water, and structure-directing agent (SDA) content were tested using a TGA spectrometer (SDT650 TGA-DSC) under airflow within 25 °C to 900 °C temperature range with a 5 °C $min^{-1}$ of heat flow rate. The DTA curve is obtained from the TGA data via differentiation using Origin software.

*The crystallinity index* (%) of the products was calculated using XRD data from 2 mg powder samples. After subtracting the baseline from the XRD spectrum, the peak areas (PA) of the most significant peaks (e.g., (501), (051), (151), (303), and (133) planes within the 22° to 25° 2θ region for TPA-Silica-1 zeolites) were calculated and summed using Origin software. Following the methodology outlined in the references,[39,40] we calculated the crystallinity index ($\alpha$) with respect to a reference that has the maximum achievable crystallinity:

$$\propto (\%) = \frac{\sum PA_{laser}}{\sum PA_{reference}} \cdot 100.$$

The crystallinity calculations from liquid samples were conducted using ATR-FTIR spectroscopy (Bruker Vertex 70 V). Peak deconvolution analyses were performed using a custom MATLAB function[41]. Prior to deconvolution, baseline correction was applied to the IR and XRD data using the asymmetric least squares algorithm implemented in Origin(Pro) software. We considered the MFI precursor (brown-filled peak) and fully-grown crystal (grey-filled peak) to calculate the IR crystallinity index. The sums were then normalized, and the crystallinity index graph shown in Figure 5C was fit to a 4-parameter logistic function

$$f(x) = d + \frac{a-d}{1+\left(\frac{x}{c}\right)^{b}},$$

where *x* denotes the time, *a* and *d* are the minimum and maximum values that can be obtained, respectively. *c* is the point of inflection (*i.e.*, the point on the S-shaped curve halfway between *a* and *d*). *b* is the Hill's slope of the curve (*i.e.*, this is related to the curve's steepness at point *c*).

*The product yield* was calculated by dividing the powder product weight by that of $SiO_2$ in the batch suspension. The measurement of the powder product was conducted post-centrifugation and drying. The calculated water and TPA contents obtained from TGA-DTA were deducted from the total mass. The batch $SiO_2$ weight was computed based on the molar formula.

**Supporting Information**

Supporting Information is available from the Wiley Online Library or from the author.

**Author Contributions**

S.G. and S.I. designed the research. S.G., M.H., and M.M.D. performed the experiments and characterizations. S.G., M.H., M.M.D., Ü.S.N. and S.I. analyzed the data. Ö.Y., G.M., M.L., and A.C. helped design the experimental setup. Ö.Y. and F.Ö.I. provided the energy calculations. B.A.K. provided facility access for some critical instruments and helped with those experiments. P.A. built the home-built ultrafast burst-mode laser. S.I. wrote the paper and supervised the study. All authors read and discussed the final manuscript.

**Acknowledgements**

This work was supported by European Research Council (ERC) under the European Union's Horizon 2020 research and innovation programme (grant agreements 853387 and 617521) and TUBITAK under projects 118F115, 120F147, and 123F473. The authors thank Dr. Paul Repgen for valuable discussions.

**Conflict of Interest Statement**

The authors declare that they have no competing interests.

**Data Availability Statement**

All data needed to evaluate the conclusions in the paper are present in the paper and/or the supplementary materials. Additional data, code, or material related to this paper may be requested from the corresponding authors.

## References

[1] Y. Li, J. Yu, *Nat. Rev. Mater.* **2021**, *6*, 1156.

[2] R. L. Siegelman, E. J. Kim, J. R. Long, *Nat. Mater.* **2021**, *20*, 1060.

[3] E. T. C. Vogt, B. M. Weckhuysen, *Chem. Soc. Rev.* **2015**, *44*, 7342.

[4] C. S. Cundy, P. A. Cox, *Chem. Rev.* **2003**, *103*, 663.

[5] W. Chaikittisilp, T. Okubo, *Science* **2021**, *374*, 257.

[6] H. Awala, J. Gilson, R. Retoux, P. Boullay, J. Goupil, V. Valtchev, S. Mintova, *Nat. Mater.* **2015**, *14*, 447.

[7] E. Ng, D. Chateigner, T. Bein, V. Valtchev, S. Mintova, *Science* **2012**, *335*, 70.

[8] Z. Liu, J. Zhu, T. Wakihara, T. Okubo, *Inorg. Chem. Front.* **2019**, *6*, 14.

[9] M. Nüchter, B. Ondruschka, W. Bonrath, A. Gum, *Green Chem.* **2004**, *6*, 128.

[10] G. A. Tompsett, W. C. Conner, K. S. Yngvesson, *Chem. Phys. Chem.* **2006**, *7*, 296.

[11] X. Chen, M. Qiu, S. Li, C. Yang, L. Shi, S. Zhou, G. Yu, L. Ge, X. Yu, Z. Liu, N. Sun, K. Zhang, H. Wang, M. Wang, L. Zhong, Y. Sun, *Angew. Chem. – Int. Ed.* **2020**, *59*, 11325.

[12] J. Chen, M. Zhang, J. Shu, M. Yuan, W. Yan, P. Bai, L. He, N. Shen, S. Gong, D. Zhang, J. Li, J. Hu, R. Li, G. Wu, Z. Chai, J. Yu, S. Wang, *Angew. Chem. – Int. Ed.* **2021**, *60*, 14858.

[13] G. Feng, P. Cheng, W. Yan, M. Borona, X. Li, J. H. Su, J. Wang, Y. Li, A. Corma, R. Xu, J. Yu, *Science* **2016**, *351*, 1188.

[14] J. P. Longtint, C.-L. Tien, *Int. J. Heat Mass Transfer* **1997**, *40*, 951.

[15] J. Noack, A. Vogel, *IEEE J. of Quant. Electron.* **1999** *35,* 1156.

[16] S. Ilday, G. Makey, G. B. Akguc, Ö. Yavuz, O. Tokel, I. Pavlov, O. Gülseren, F. Ömer Ilday, *Nat. Commun.* **2017**, *8*.

[17] G. Makey, S. Galioglu, R. Ghaffari, E. D. Engin, G. Yıldırım, Ö. Yavuz, O. Bektaş, S. Nizam, Ö. Akbulut, Ö. Şahin, K. Güngör, D. Dede, H. V. Demir, F. Ö. Ilday, S. Ilday, *Nat. Phys.* **2020**, *16*, 795.

[18] K. Eidmann, J. Meyer-Ter-Vehn, T. Schlegel, S. Hü, *Phys. Rev. E.* **2000**, *62*, 1202.

[19] A. L. Gaeta, *Phys. Rev. Lett.* **2000**, *84*, 3582.

[20] D. Zhang, B. Gökce, S. Barcikowski, *Chem. Rev.* **2017**, *117*, 3990.

[21] M. Navarro, Ú. Mayoral, E. Mateo, R. Lahoz, G. F. De La Fuente, J. Coronas, *Chem. Phys. Chem.* **2012**, *13*, 736.

[22] X. Li, Z. Wang, J. Zheng, S. Shao, Y. Wang, Y. Yan, *Chin. J. Catal.* **2011**, *32*, 217.

[23] Q. Ge, J. Shao, Z. Wang, Y. Yan, *Micropor. Mesopor. Mater.* **2012**, *151*, 303.

[24] M. Mazur, P. S. Wheatley, M. Navarro, W. J. Roth, M. Položij, A. Mayoral, P. Eliášová, P. Nachtigall, J. Čejka, R. E. Morris, *Nat. Chem.* **2016**, *8*, 58.

[25] H. Dai, Y. Shen, T. Yang, C. Lee, D. Fu, A. Agarwal, T. T. Le, M. Tsapatsis, J. C. Palmer, B. M. Weckhuysen, P. J. Dauenhauer, X. Zou, J. D. Rimer, *Nat. Mater.* **2020**, *19*, 1074.

[26] M. Thommes, K. Kaneko, A. V. Neimark, J. P. Olivier, F. Rodriguez-Reinoso, J. Rouquerol, K. S. W. Sing, *Pure Appl. Chem.* **2015**, *87*, 1051.

[27] W. Corkery, B. W. Ninham, *Zeolites* **1997**, *18*, 379.

[28] M. Tawalbeh, F. H. Tezel, B. Kruczek, S. Letaief, C. Detellier, *J. Porous Mater.* **2013**, *20*, 1407.

[29] M. Razavian, S. Fatemi, M. Masoudi-Nejad, *Adsorpt. Sci. Technol.* **2014**, *32*, 73.

[30] L. Tosheva, B. Mihailova, V. Valtchev, J. Sterte, *Micropor. Mesopor. Mater.* **2000**, *39*, 91.

[31] D. Lesthaeghe, P. Vansteenkiste, T. Verstraelen, A. Ghysels, C. E. A. Kirschhock, J. A. Martens, V. Van Speybroeck, M. Waroquier, *J. Phys. Chem. C* **2008**, *112*, 9186.

[32] C. Y. Hsu, A. S. T. Chiang, R. Selvin, R. W. Thompson, *J. Phys. Chem. B* **2005**, *109*, 18804.

[33] C. E. A. Kirschhock, R. Ravishankar, F. Verspeurt, P. J. Grobet, P. A. Jacobs, J. A. Martens, *J. Phys. Chem. B* **1999**, *103*, 4965.

[34] P. Elahi, C. Kerse, H. Hoogland, Ö. Akçaalan, D. K. Kesim, B. Çetin, R. Holzwarth, F. Ö. Ilday, B. Öktem, M. D. Aşık, S. Yavaş, H. Kalaycıoğlu, *Nature* **2016**, *537*, 84.

[35] J. Q. Krimmer, “Quiver magnitude dependent color in 2D and 3D, MATLAB Central File Exchange,” **2023**.

[36] J. Hedlund, S. Mintova, J. Sterte, *Micropor. Mesopor. Mater.* **1999**, *28*, 185.

[37] B. J. Schoeman, J. Sterte, J. E. Otterstedt, *Stud. Surf. Sci. Catal.* **1994**, *83*, 49.

[38] Y. Zhu, Z. Hua, Y. Song, W. Wu, X. Zhou, J. Zhou, J. Shi, *J. Catal.* **2013**, *299*, 20.

[39] J. Qi, T. Zhao, X. Xu, F. Li, G. Sun, *J. Porous Mater.* **2011**, *18*, 509.

[40] K. Jiao, X. Xu, Z. Lv, J. Song, M. He, H. Gies, *Micropor. Mesopor. Mater.* **2016**, *225*, 98.

[41] T. O’Haver, “MATLAB Central File Exchange” can be found under https://www.mathworks.com/matlabcentral/fileexchange/23452-ipf-arg1-arg2-arg3-arg4, **2024**.

# Supporting Information

## Ultrafast laser synthesis of zeolites

*Sezin Galioglu[1,*,†], Mehdi Hagverdiyev[1,*], Meryem M. Doğan[1], Özgün Yavuz[2], Ü. Seleme Nizam[2], Ghaith Makey[1], Aladin Şura[2], Mesut Laçin[2], Burcu Akata Kurç[4], Parviz Elahi[5], F. Ömer Ilday[2,3], Serim Ilday[2,3,†]*

[1] UNAM-National Nanotechnology Research Center, Bilkent University, Ankara, 06800, Türkiye
[2] Faculty of Electrical Engineering and Information Technology, Ruhr University Bochum, 44801, Germany
[3] Faculty of Physics and Astronomy, Ruhr University Bochum, 44801, Germany
[4] Micro and Nanotechnology Programme, Middle East Technical University, 06800, Ankara, Türkiye
[5] Faculty of Engineering, Özyeğin University, 34794, İstanbul, Türkiye

*The authors have equally contributed to this study.
†To whom correspondence should be addressed: sezin@unam.bilkent.edu.tr & serim.ilday@rub.de

**Video Captions**

**Video 1.** Real-time video recording of an experiment demonstrating the color change of the precursor solution from transparent to opaque white.

**Video 2**. Real-time video recordings (2$^{nd}$ row) and vector field analyses of flow patterns (3$^{rd}$ row) for three experiments (1$^{st}$ row), where the laser beam is focused at three different locations: the glass-liquid interface (left), a position near the interface (middle), and closer to the center of the bulk liquid (right).

**Video 3.** Real-time video recording of an experiment where the laser beam is positioned away from the glass-liquid interface so that it does not promote multiphoton absorption. As a result, there are no laser-induced high-speed convective flows, bubble formation, or plasma generation. Instead, the solution is agitated using a magnetic stirrer.

**Video 4.** Real-time video recording of the experiment where the laser synthesis is halted and resumed multiple times.

## Estimation of the energy absorption

We begin by measuring the transmitted power and calculating the optical reflection and absorption losses of the laser beam at the various optical interfaces, from which we determine the energy deposited in the liquid (**Figure S1**).

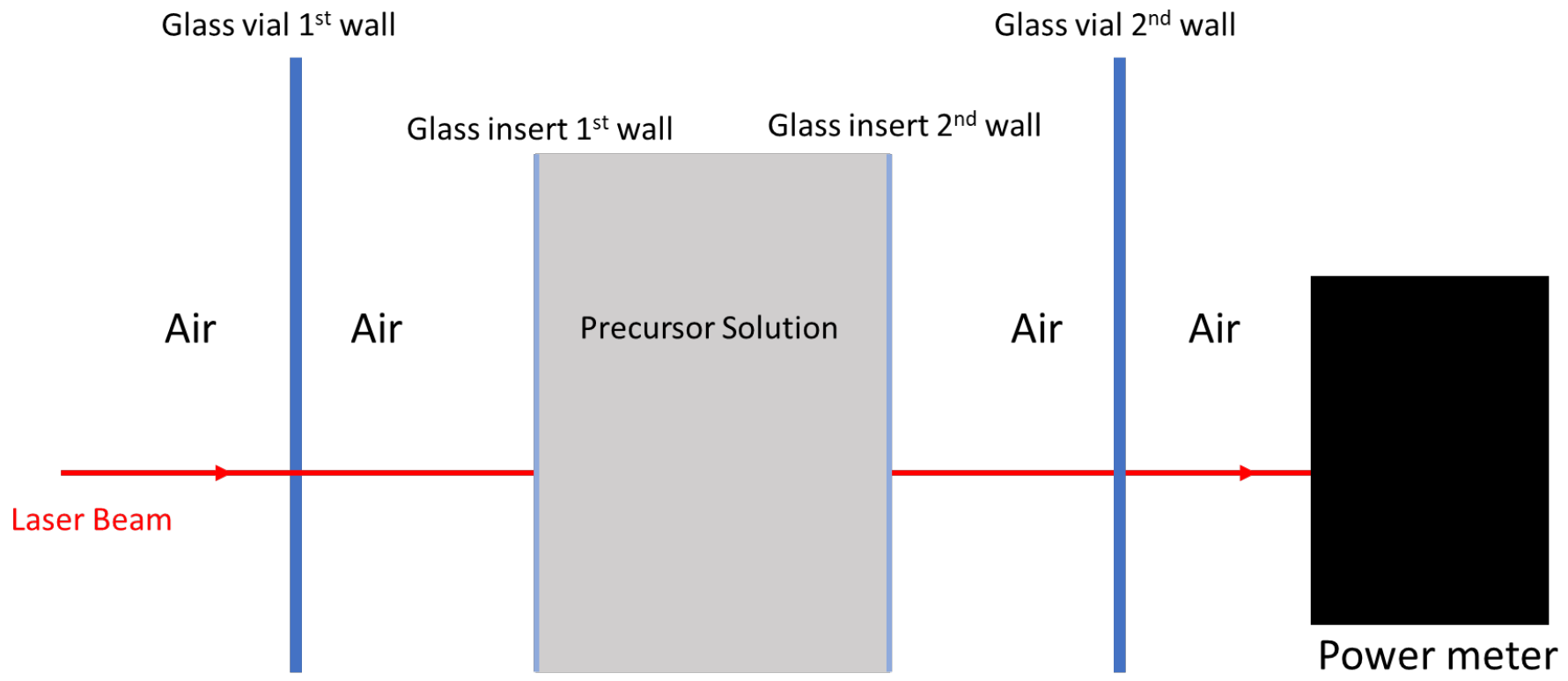

**Figure S1.** Schematic illustrating the primary optical loss mechanisms resulting from multiple interfaces and media.

**Figure S2** shows the interfaces where we measured light transmittance from the air–glass vial–air ($T_1$ and $T_2$), air–glass insert–air ($T_3$ and $T_4$), and glass insert–precursor solution–glass insert ($T_5$ and $T_6$) interfaces. We calculated $T_1$ and $T_2$ by measuring the light passing through an empty glass vial. Similarly, $T_3$ and $T_4$ were measured using an empty glass insert. The glass insert and vial wall thicknesses were measured as 0.92 mm and 1.16 mm.

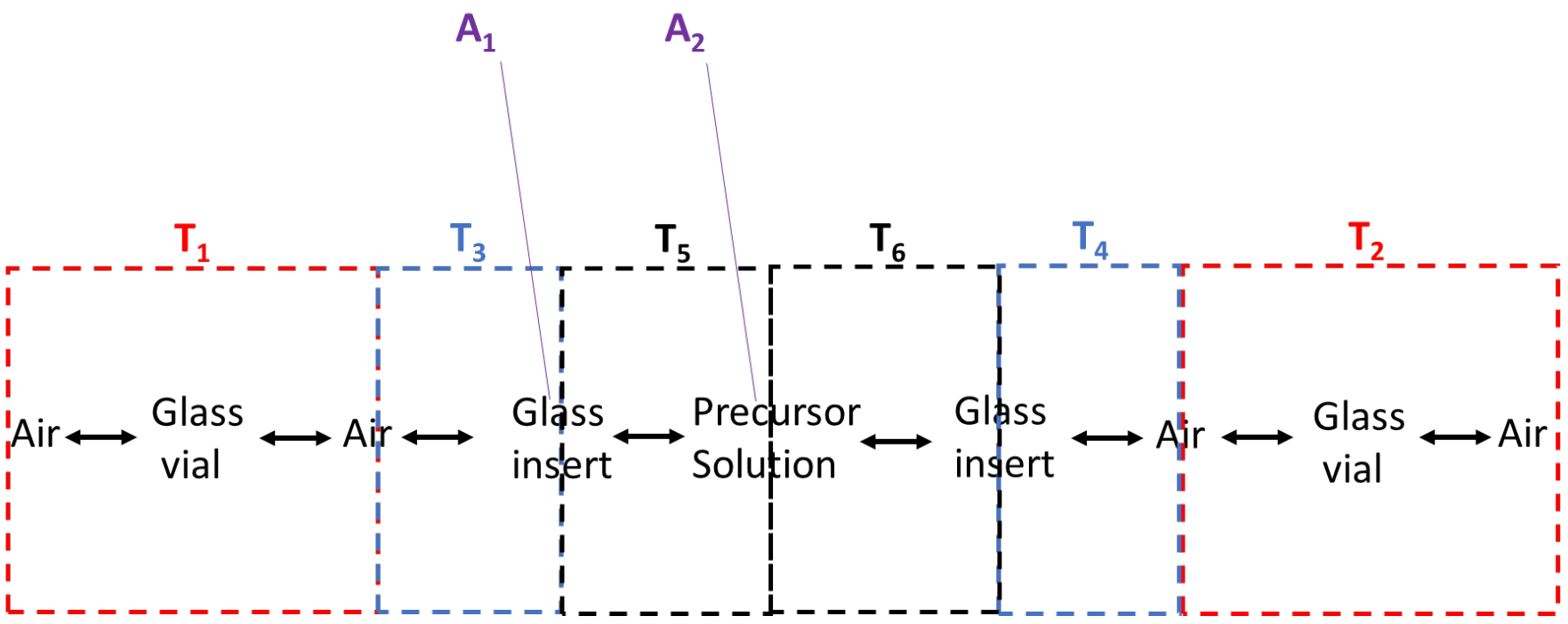

**Figure S2.** Schematic illustrating the types of measurements taken at each interface.

Next, the losses due to reflection and absorption by the 1st wall of the glass insert (A1) and the precursor solution (A2) were measured by filling the glass insert with the precursor solution. During absorbance measurements, the transmitted power naturally fluctuated due to the formation of plasma, fluid flows, and bubbles. Therefore, we measured the average value. Furthermore, we assumed that the refractive index of the precursor solution was similar to that of the water. Based on the experimental measurements, we determined that 22.5% of the incident average laser power is absorbed within the precursor solution.

**Calculation of the thermal gradients**

We will next estimate the magnitude of the thermal gradients formed around the beam focus. Due to the presence of multiple nonlinear dissipative and scattering mechanisms, such as multi-photon absorption, plasma and cavitation bubble formation, and convective flows, an accurate calculation is prohibitively difficult. However, using the measured absorbed power and pulse parameters, and aided by a set of reasonable approximations, we can calculate an approximate value, likely a lower bound.

Given the empirically determined absorption of 22.5% of the laser power in the solution, we first calculate the time it should take to heat the entire liquid from room temperature of 21 °C to 95 °C, the steady state average temperature, $\Delta T_{\text{ave}}$, measured directly in the experiments. For this calculation, we assume the liquid to be pure water, and we neglect other dissipative processes, such as nucleation and crystal growth, bubble generation, plasma formation, and convective flows. For an incident laser power of 6 W, the absorbed power is, $P_{\text{abs}} = 1.35$ W. Then, the time to raise the temperature of the bulk liquid to its steady state value is given by

$$\Delta t_{\text{bulk}} = \frac{m_{\text{tot}} C \Delta T_{\text{ave}}}{P_{\text{absorbed}}} \cong 90\ s,$$

where $\Delta T_{\text{ave}}$ is 74 °C, $C = 4.186$ J g$^{-1}$ °C$^{-1}$ is the specific heat of the water, and $m_{\text{tot}} = 0.4$ g is the mass of the liquid, assumed to be water, for its volume of $V_{\text{total}} = 400$ mm$^3$. The calculated time of approximately 90 s for the bulk liquid to reach the experimentally measured temperature is in excellent agreement with the measured time shown in Figure 1D.

Now that we have determined the power absorbed and corresponding temperature rise on average for the entire liquid, next, we focus on the more challenging problem of estimating the same values for the laser-liquid interaction volume.

The beam is focussed to an ellipsoid with a volume ($V_{\mathrm{focal}}$), which can be calculated using the Rayleigh length ($z_{\mathrm{R}}$) and beam waist ($\omega_0$) as below. The beam spot size was measured to be 9 µm using a beam profiler (Thorlabs BP209-IR2/M).

$$V_{\mathrm{focal}} = \frac{4}{3}\pi {\omega_0}^2 z_{\mathrm{R}},$$

$$z_{\mathrm{R}} = \frac{\pi {\omega_0}^2}{\lambda},$$

$$z_{\mathrm{R}} = \frac{(3.14)(4.5\ x\ 10^{-6})^2}{(1040\ x\ 10^{-9})} = 60\ \mu\mathrm{m},$$

$$V_{\mathrm{focal}} = \frac{4}{3}(3.14)(4.5\ \mu m)^2(60\ \mu m) = 5.1 \cdot 10^{-6}\ \mathrm{mm}^3,$$

$$V_{\mathrm{focal}} = 5100\ \mu\mathrm{m}^3.$$

The liquid is almost completely transparent at the beginning of the synthesis, which means there is negligible linear absorption. Even later, when it turns opaque white, this indicates increased scattering, not absorption. Consistent with the liquid being largely water, we assume linear and even two-photon absorption at the laser wavelength of 1 µm to be negligible, and that the dominant absorption mechanism, at least prior to the formation of a plasma, is three-photon absorption. Because three-photon absorption scales with the square of the peak intensity, the absorbed power for a Gaussian pulse shape and beam shape that is also Gaussian will be approximately proportional to the cube of a Gaussian ellipsoid. Integrating over such a shape, compared to linear absorption of the same Gaussian ellipsoid, the effective volume is decreased by a factor of $3^{3/2} \cong 5.2$. Thus, the interaction volume, $V_{\mathrm{int}}$, over which appreciable absorption occurs, is approximately one-fifth of the focal volume, or $V_{\mathrm{int}} = 980$ µm$^3$. Even after the establishment of a plasma, which introduces linear absorption, the interaction volume is not expected to change substantially because the plasma will remain largely confined to it.

Now, we can calculate the temperature rise caused by the absorption of a single pulse. We assume three-photon absorption is dominant, thus there is negligible absorption outside of the interaction volume. We momentarily neglect all dissipative effects occurring during the pulse (within 300 fs), as well as higher-order multi-photon processes. Also, as before, we ignore all the chemicals, and proceed as if the liquid is pure water. We further ignore any phase changes, such as boiling, which need a far longer timescale, during the ultrashort pulse. For the absorbed power of $P_{\mathrm{abs}} = 1.35$ W and pulse repetition rate of $f_{\mathrm{R}} = 200$ kHz, the absorbed pulse energy is $E_{\mathrm{abs}} = P_{\mathrm{abs}} f_{\mathrm{R}}^{-1} = 6.75$ μJ, which becomes the heat source, $Q$. The temperature rise due to a single pulse is

$$\Delta T_{\mathrm{pulse}} = \frac{Q}{m_{\mathrm{int}} C} \cong 1600°\mathrm{C}\ (\mathrm{or\ K}),$$

where $\Delta T_{\mathrm{pulse}}$ is the heat increment per pulse, $C = 4.186$ J $g^{-1}$ $°C^{-1}$ is the specific heat of the water, and $m_{\mathrm{int}} = 980$ pg is the water mass within the interaction volume. We further note that nearly all the energy is first coupled to the electronics, which equilibrate by transferring the energy to their atoms within several 100 fs. This peak temperature is reached only within approximately half of the beam diameter or about 2 μm due to three-photon absorption, whereas there is negligible absorption of laser light laterally further away and the temperature remains at its average value. Therefore, the temperature gradient reaches $\sim 10^6$ K $mm^{-1}$ albeit momentarily, because the interaction volume begins to cool rapidly immediately at the end of the pulse until the arrival of the next pulse within 5 μs (for a repetition rate of 200 kHz) due to various dissipative effects (Figure 1C). These effects include mechanical energy associated with the convective flows (along the axial direction of the beam), plasma and cavitation bubble formation, pressure waves, blackbody radiation, and the chemical reactions. With subsequent pulses, the peak temperatures builds up further, even though cooling will also be faster. Based on other ultrafast light-matter interactions, it is likely that peak temperatures achieved momentarily after subsequent pulses can rise as much as 8,000 K. This value is consistent with the blackbody temperature corresponding to the experimentally observed blue-purple light emission. Consequently, we conclude the peak spatial thermal gradients calculated above may constitute a lower bound.

Finally, we will estimate the transit time of any chemical entering and exiting the interaction volume, *i.e.*, the tiny ultrafast reactor. This task is complicated by the difficulty to calculate the flow speeds achieved within the focal volume. The bubbles, which vary in size, are hard to monitor accurately due to imaging and tracking limitations, and their movement driven by buoyancy forces further complicates the situation. The interactions between the flows and bubbles cause light scattering, while plasma formation significantly alters the absorption dynamics. Experimentally, we only have a lower bound for the flow speeds far from the interaction region to be 1 mm $s^{-1}$. However, the flow cross-section has expanded to 10s of micron, even 100 μm, where the lower boundary was established by tracking bubbles. At the center of the interaction volume, where the flow must be the fastest, the corresponding diameter is about 5 μm, which suggests that the flow can be in the order of 1 m $s^{-1}$ or even higher by the continuity equation for an incompressible fluid. While this is a highly uncertain estimate, such high speeds are consistent with the observed experiment durations for the zeolite synthesises to be largely completed because this requires nearly all of the liquid volume to pass through the interaction volume at least once. Assuming top speeds of 1 m $s^{-1}$, the time it takes to cross the length of the three-photon interaction volume, ~50 μm, is 50 μs, during which 10 pulses would be incident. Given the uncertainty in the assessment of the fluid speeds, this estimate should be taken as an order-of-magnitude assessment.

**Examples of laser-synthesized zeolites**

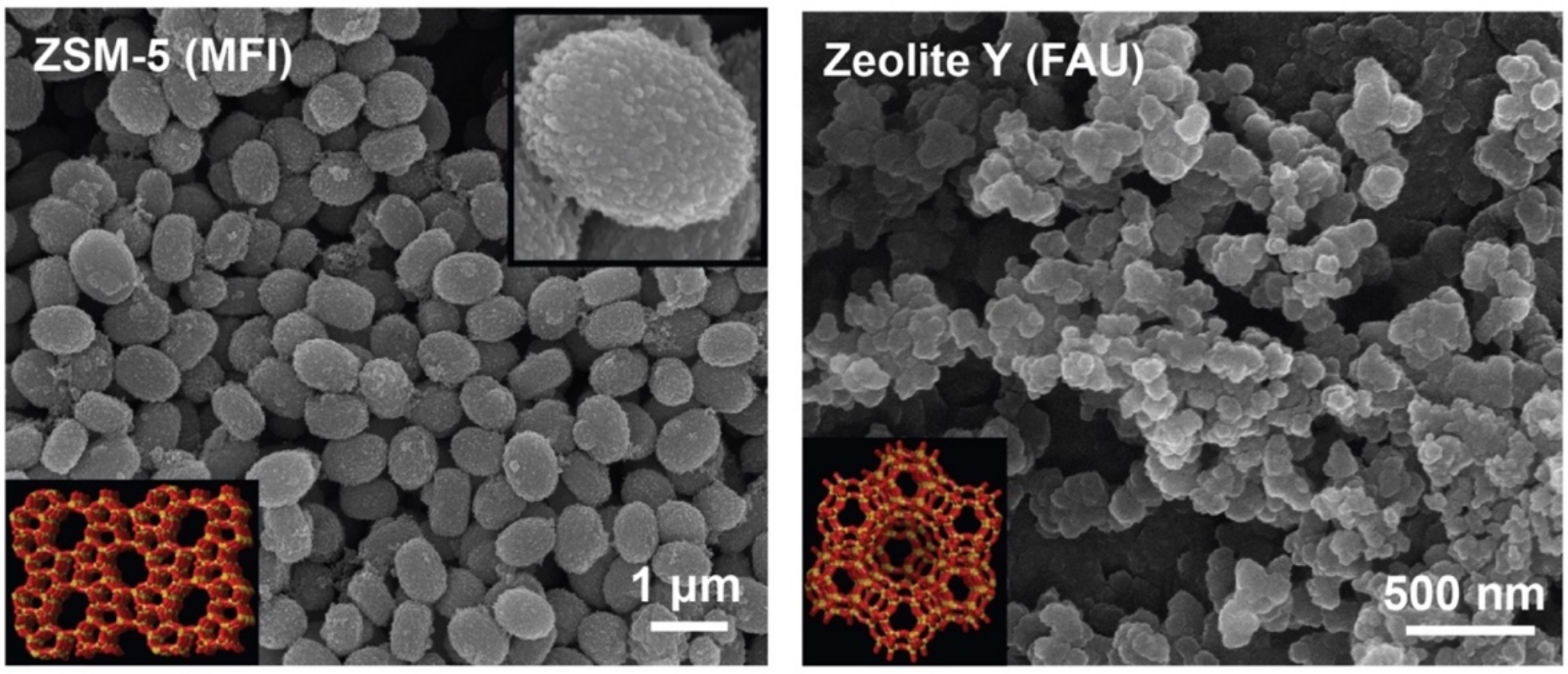

**Figure S3.** SEM images of (left) microporous ZSM-5 (MFI type) and (right) template-free Zeolite Y (FAU type) synthesized via laser synthesis method. Insets showing unit cell structures.

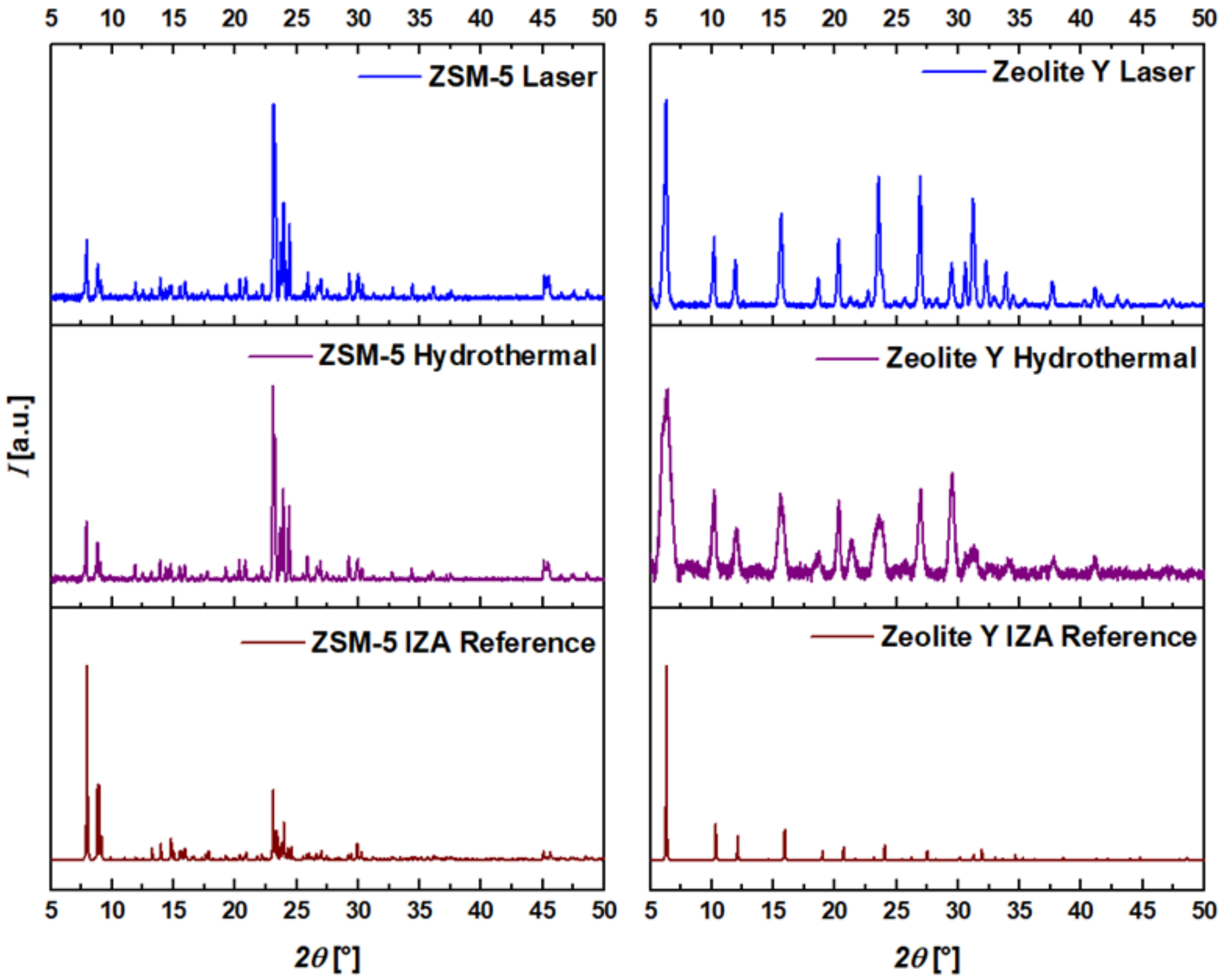

**Figure S4.** XRD patterns of microporous (left) ZSM-5 and (right) Zeolite Y crystals synthesized via laser and hydrothermal synthesis methods. Reference XRD patterns are obtained from the International Zeolite Association (IZA)'s webpage.

**Repeatability of the laser-synthesized zeolites**

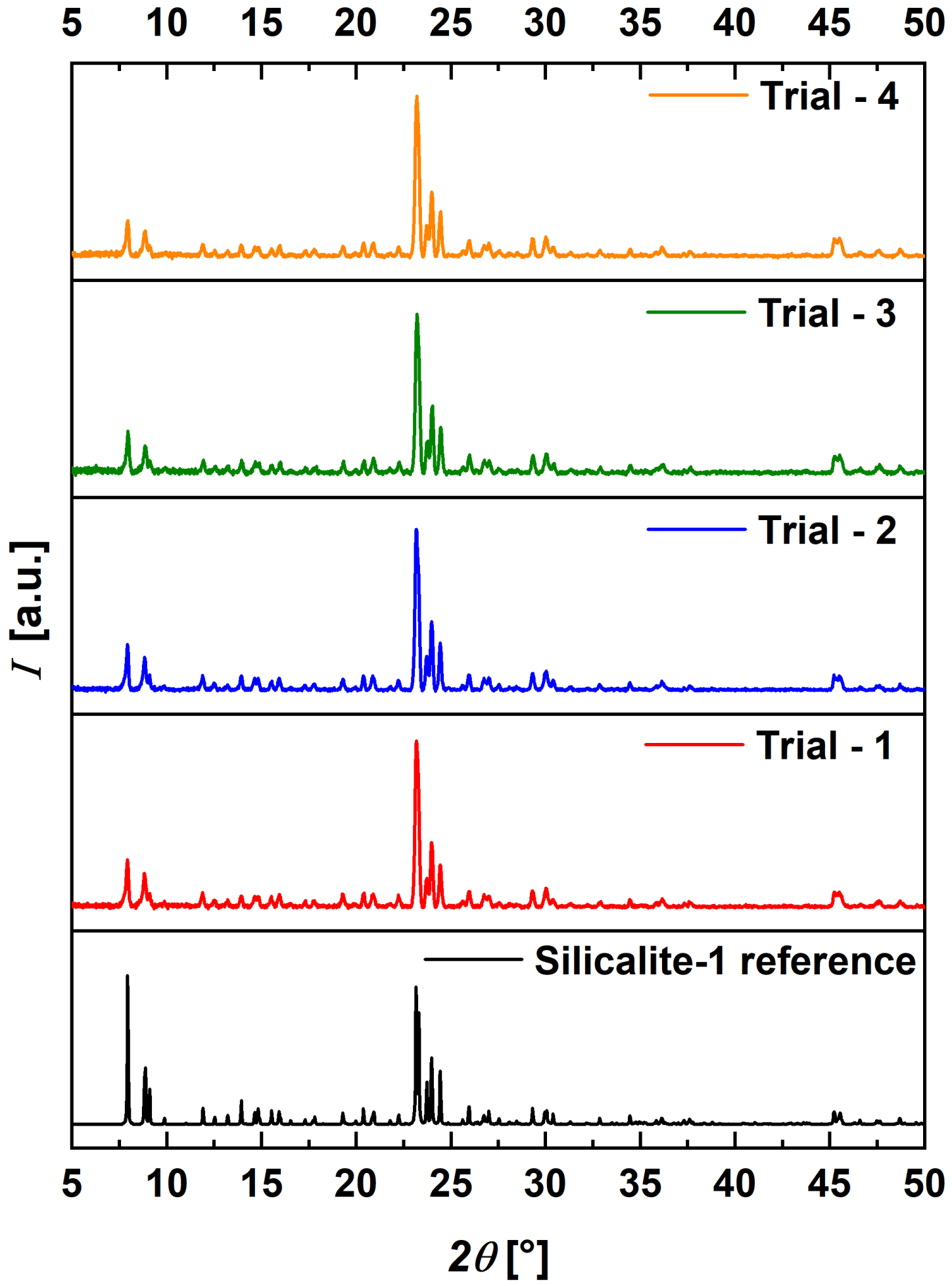

**Figure S5.** XRD patterns of laser-synthesized TPA-silicalite-1zeolites, where the original experiment (Trial 1 with 72 % crystallinity) was repeated after 2 days (Trial 2 with 86 % crystallinity), 1 week (Trial 3 with 75 % crystallinity), and 1 month (Trial 4 with 80 % crystallinity). They are compared to the reference XRD pattern approved by the International Zeolite Association (IZA).

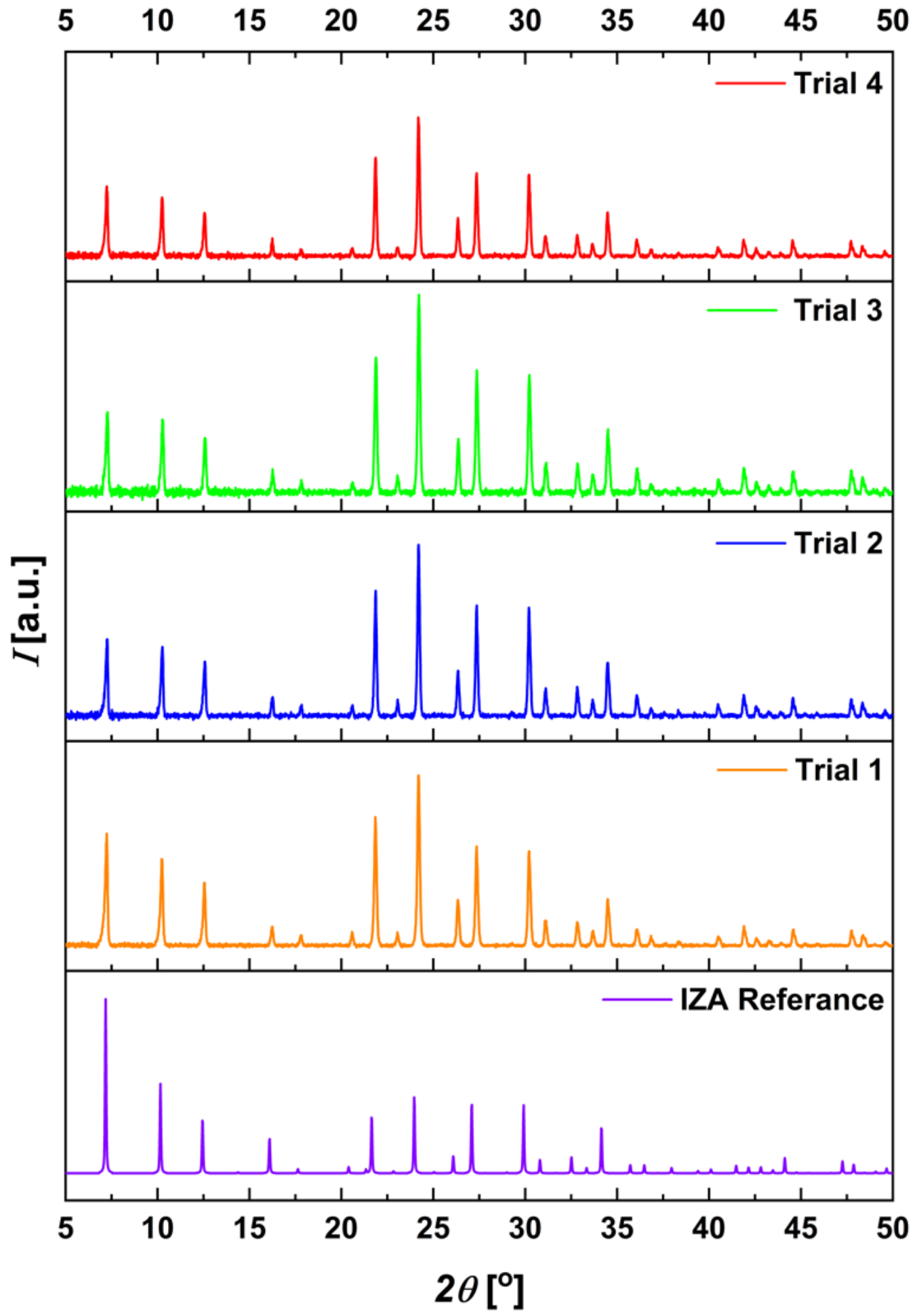

**Figure S6.** XRD patterns of laser-synthesized Zeolite A, where the original experiment (Trial 1 with 100 % crystallinity) was repeated after 1 week (Trial 2 with 97.5 % crystallinity), 1 month (Trial 3 with 93 % crystallinity) and 3 months (Trial 4 with 99 % crystallinity). They are compared to the reference XRD pattern approved by the International Zeolite Association (IZA).

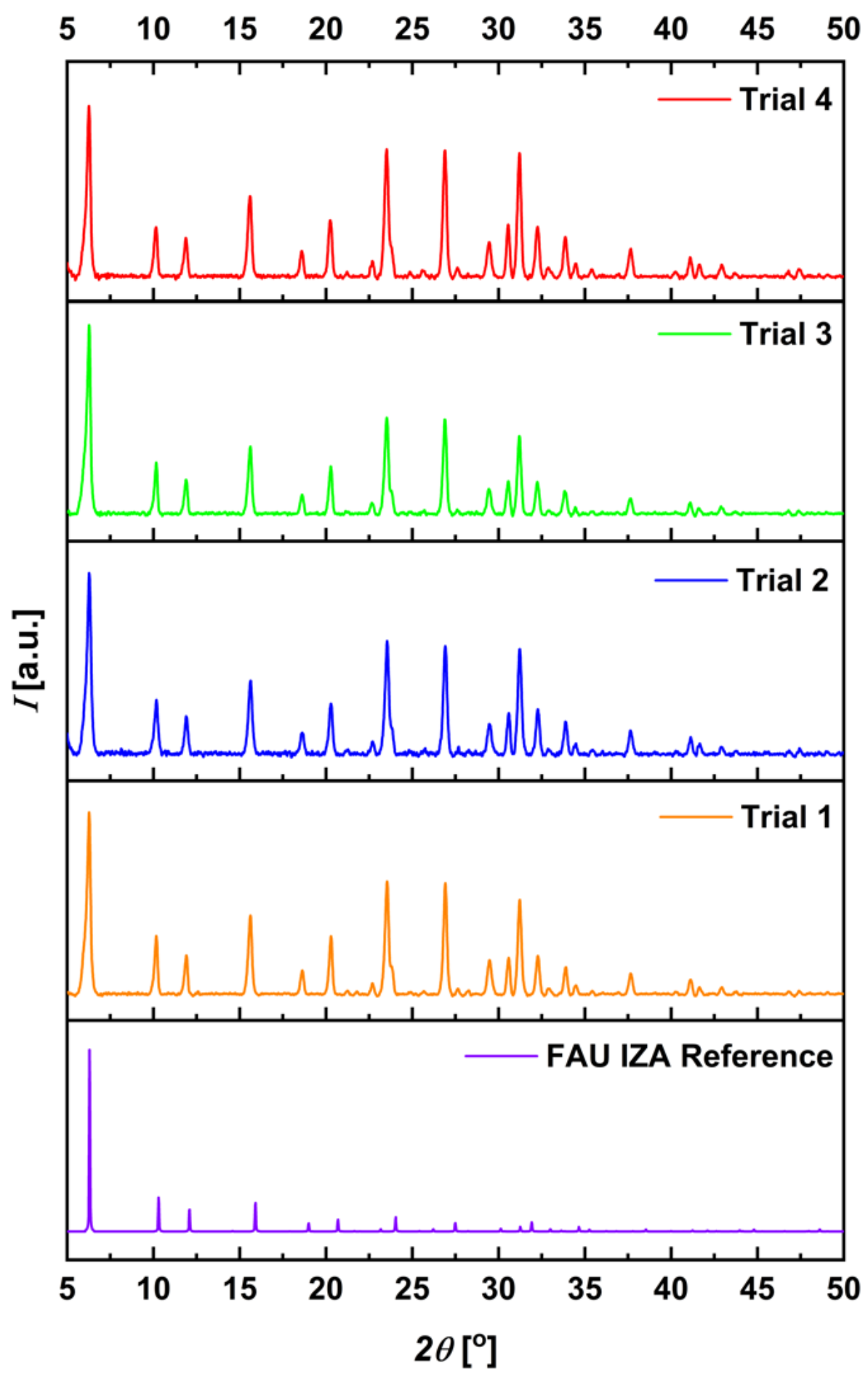

**Figure S7.** XRD patterns of laser-synthesized Zeolite Y, where the original experiment (Trial 1 with 96.4 % crystallinity) was repeated after 4 months (Trial 2 with 92.1 % crystallinity), 7 months (Trial 3 with 87.2 % crystallinity), and 8 months (Trial 4 with 100 % crystallinity). They are compared to the reference XRD pattern approved by the International Zeolite Association (IZA).

**Comparison of laser- and hydrothermal-synthesized TPA-Silicalite-1 zeolites**

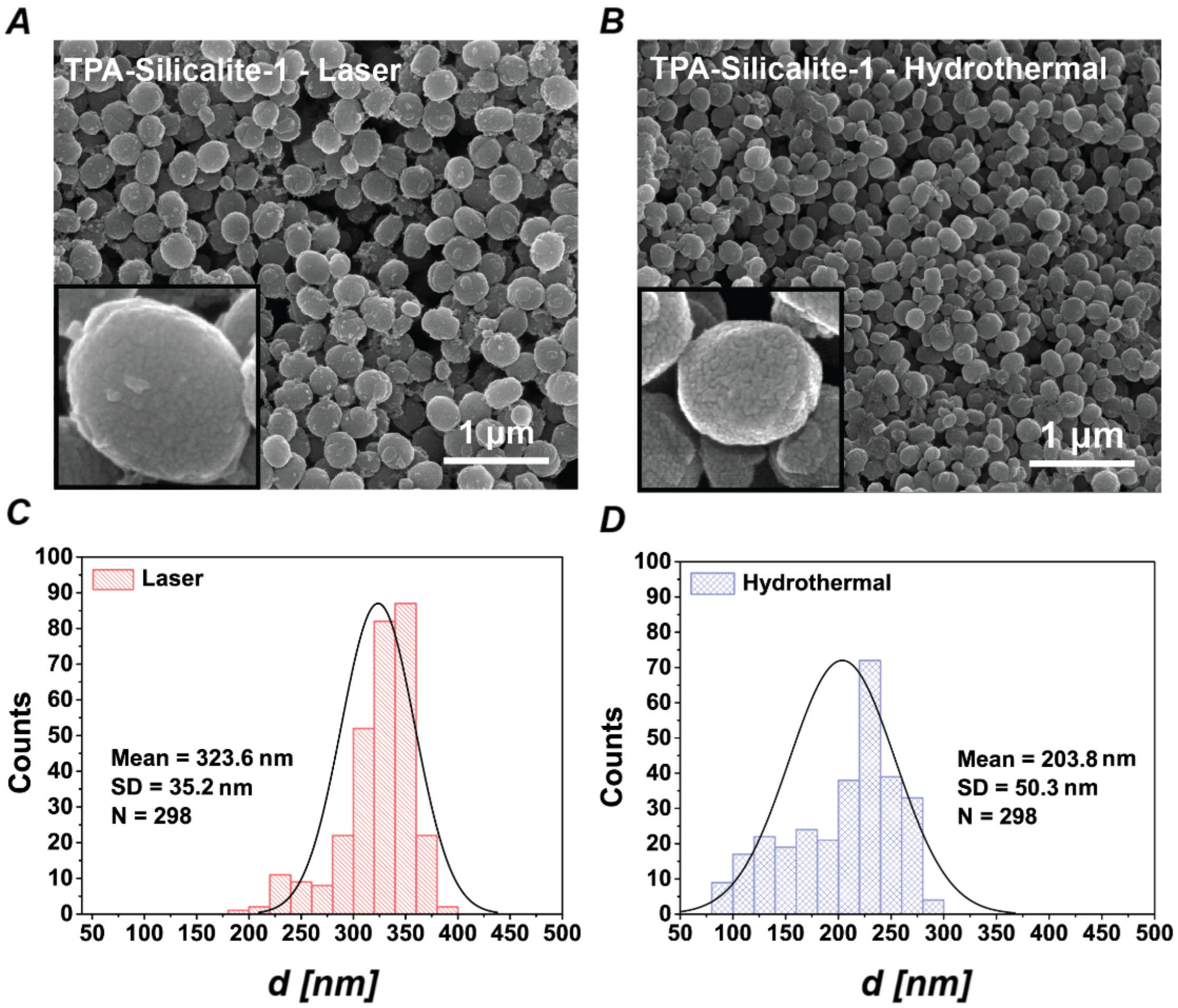

**Figure S8.** SEM images and the plots show the comparison of the particle size distribution for **(A,C)** laser- and **(B, D)** hydrothermal-synthesized zeolites.

**Reducing the average crystal size of laser-synthesized TPA-Silicalite-1 zeolites**

In previous hydrothermal synthesis trials where the effect of water content within the precursor suspension was studied, the linear growth rate of zeolite crystals seemed to increase with more water.[1] Increasing the water content, all else constant, means a decrease in silica concentration and alkalinity in the synthesis mixture. Alkalinity affects number of particles and their linear growth rate (*i.e.* size). Higher alkalinity induces more number of nuclei to grow. As alkalinity increases (*i.e.,* water content decreases), particles tend to grow slower, resulting in narrower size distribution for hydrothermal and laser synthesis methods.

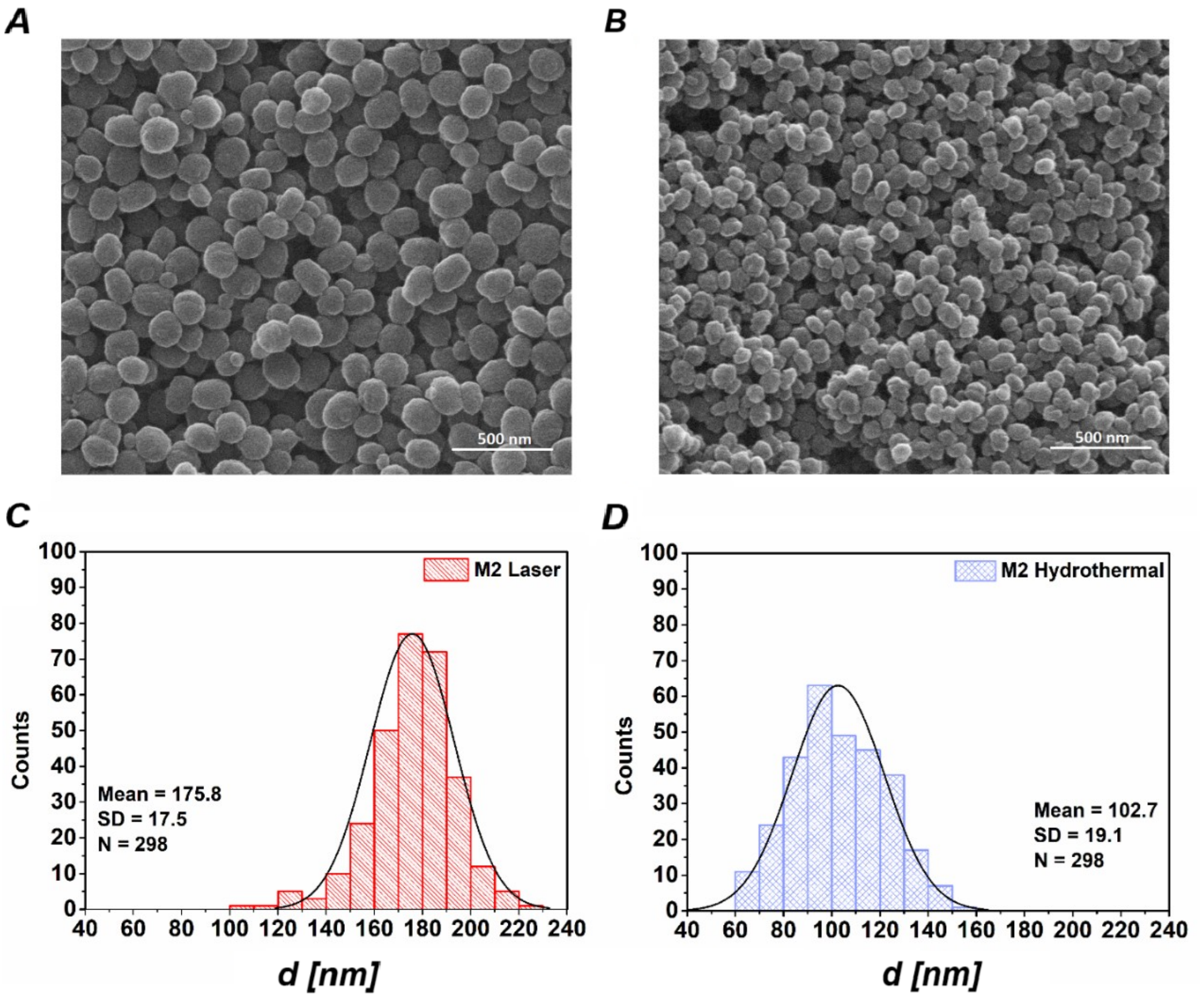

**Figure S9.** Comparison of **(A)** laser- and **(B)** hydrothermal-synthesized TPA – Silicalite-1 crystals using M2 (M2: 25 $SiO_2$: 9 TPAOH: 480 $H_2O$: 100 EtOH) molar formula. The rows above and below show SEM images and particle size distributions.

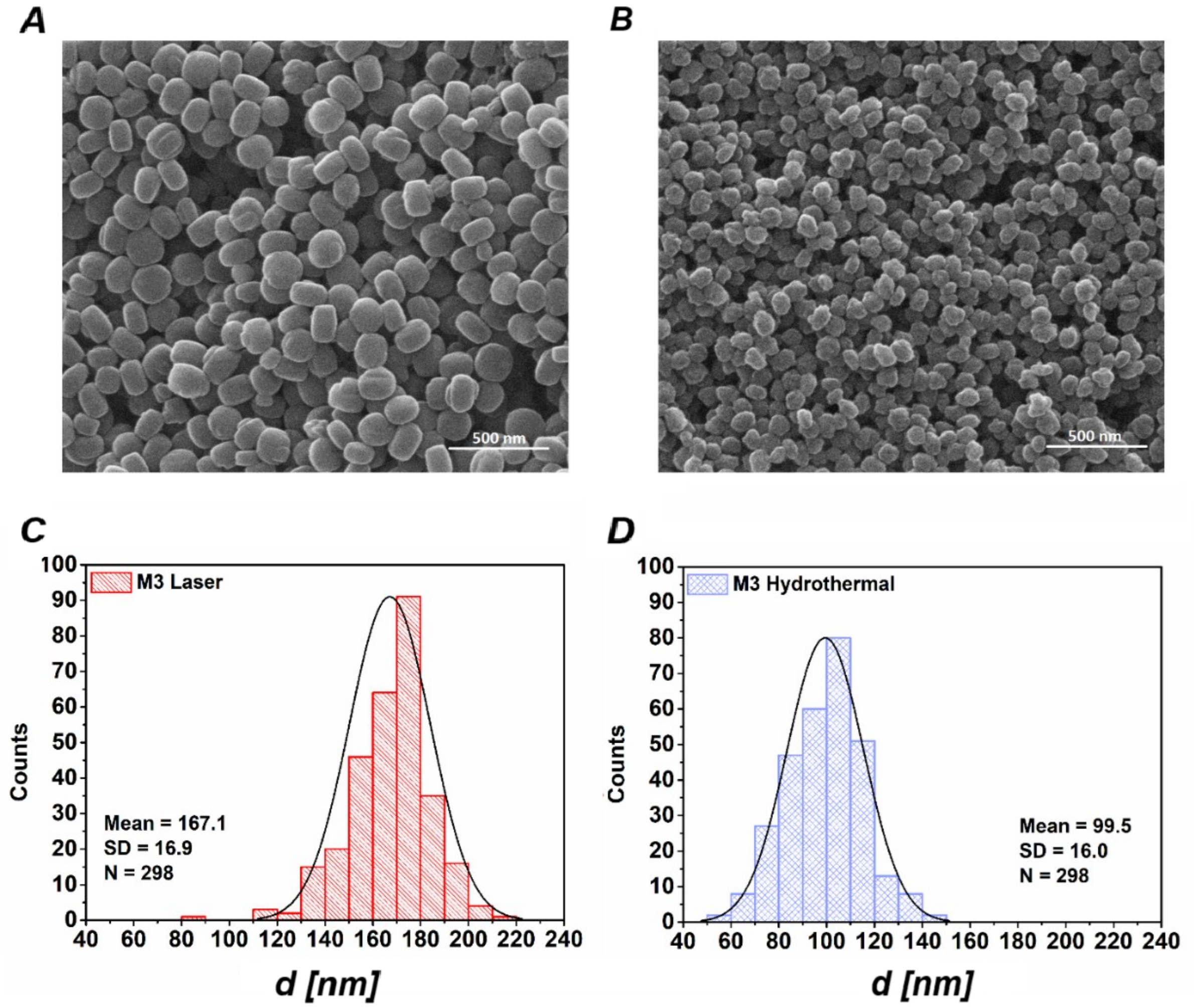

**Figure S10.** Comparison of **(a)** laser- and **(b)** hydrothermal-synthesized TPA – Silicalite-1 crystals using M3 (M3: 25 $SiO_2$: 9 TPAOH: 450 $H_2O$: 100 EtOH) molar formula. The rows above and below show SEM images and particle size distributions.

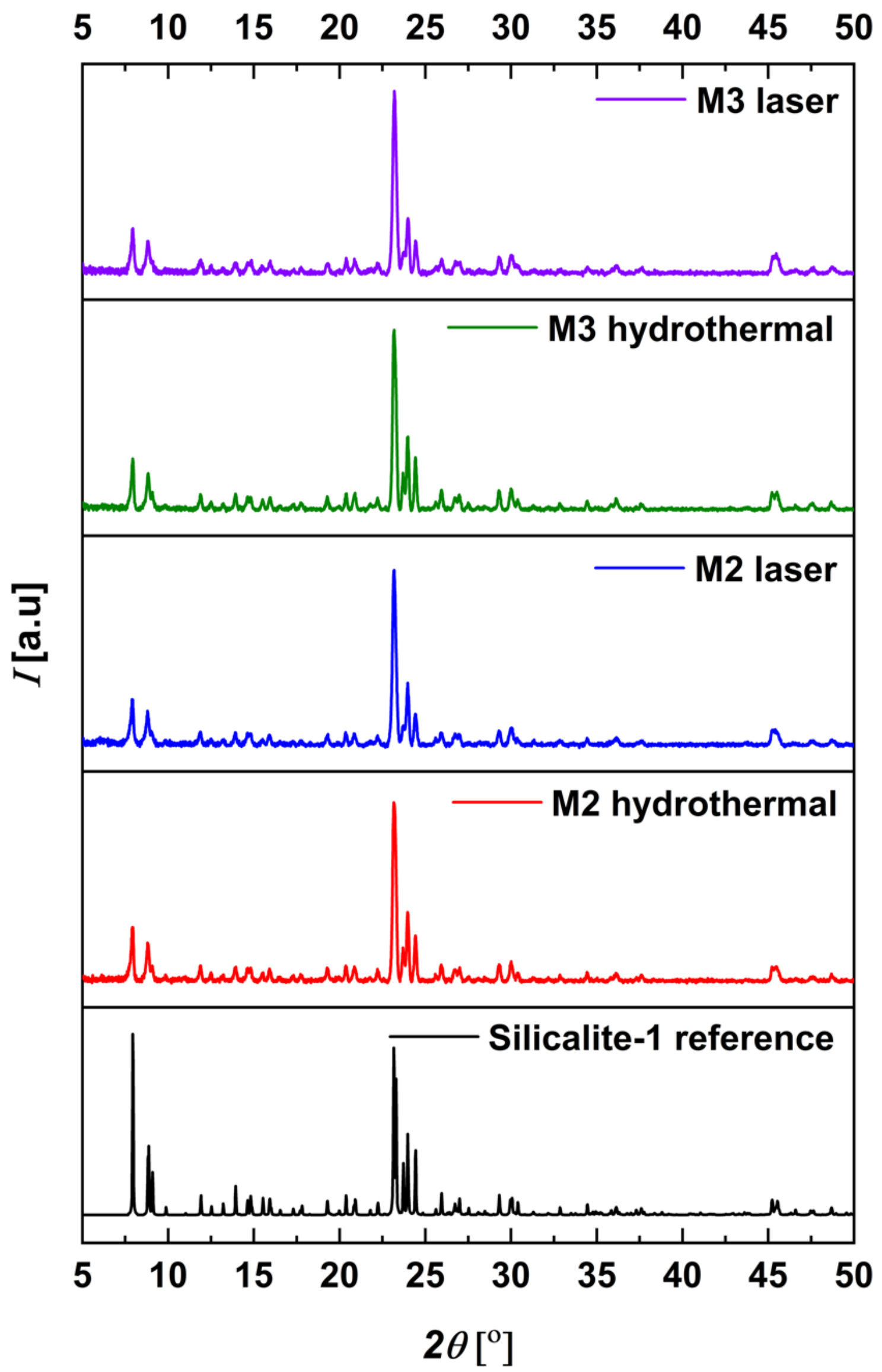

**Figure S11.** XRD spectrum of laser- and hydrothermal-synthesized TPA-Silicalite-1 zeolites using M2 and M3 molar formulas and compared to the IZA reference.

**Thermogravimetric (TGA) and Differential Thermal (DTA) analysis**

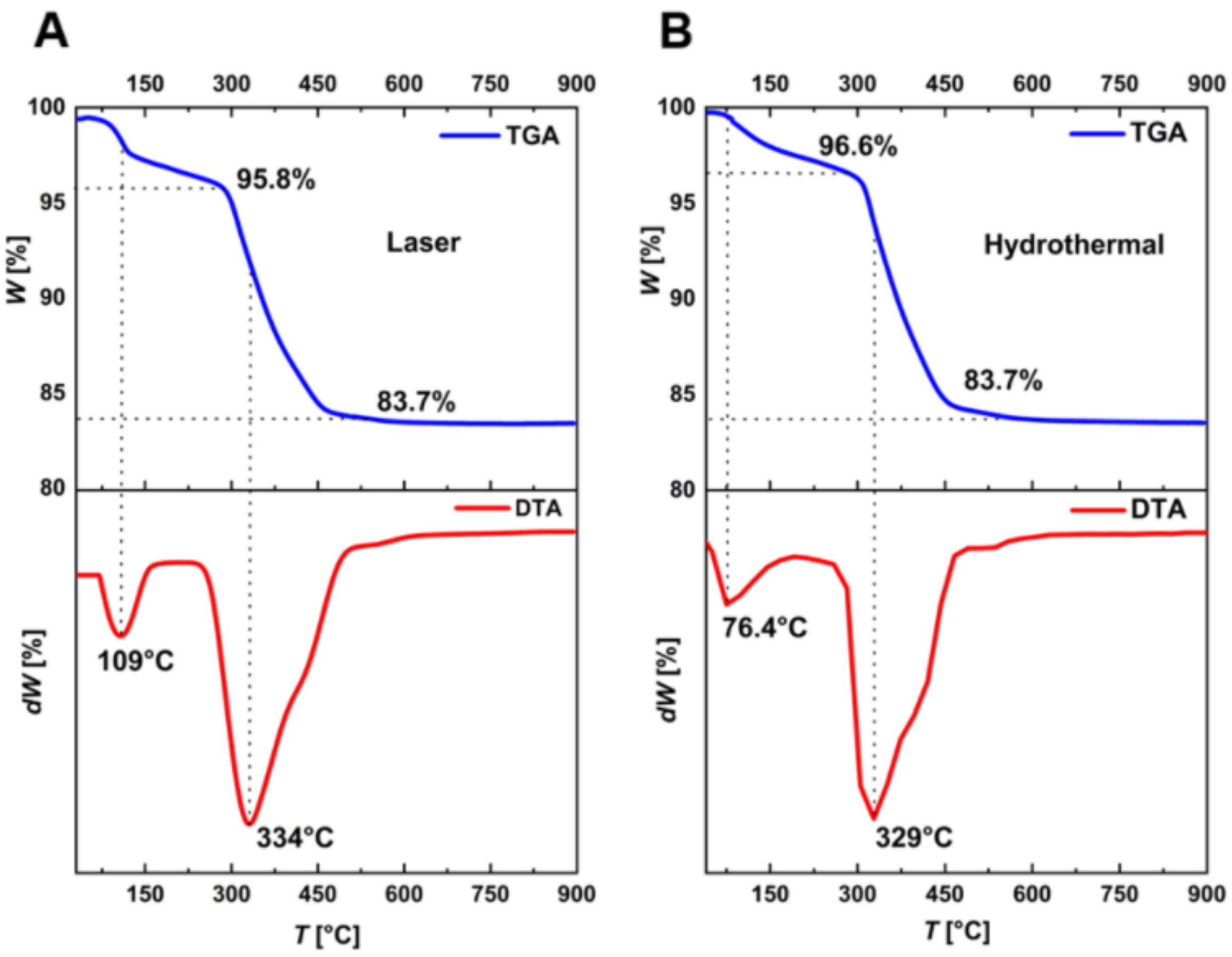

**Figure S12.** TGA and DTA curves of TPA-Silicalite-1 synthesized via **(A)** laser and **(B)** hydrothermal methods

## Brunauer-Emmett-Teller (BET) and Density Functional Theory (DFT) analyses

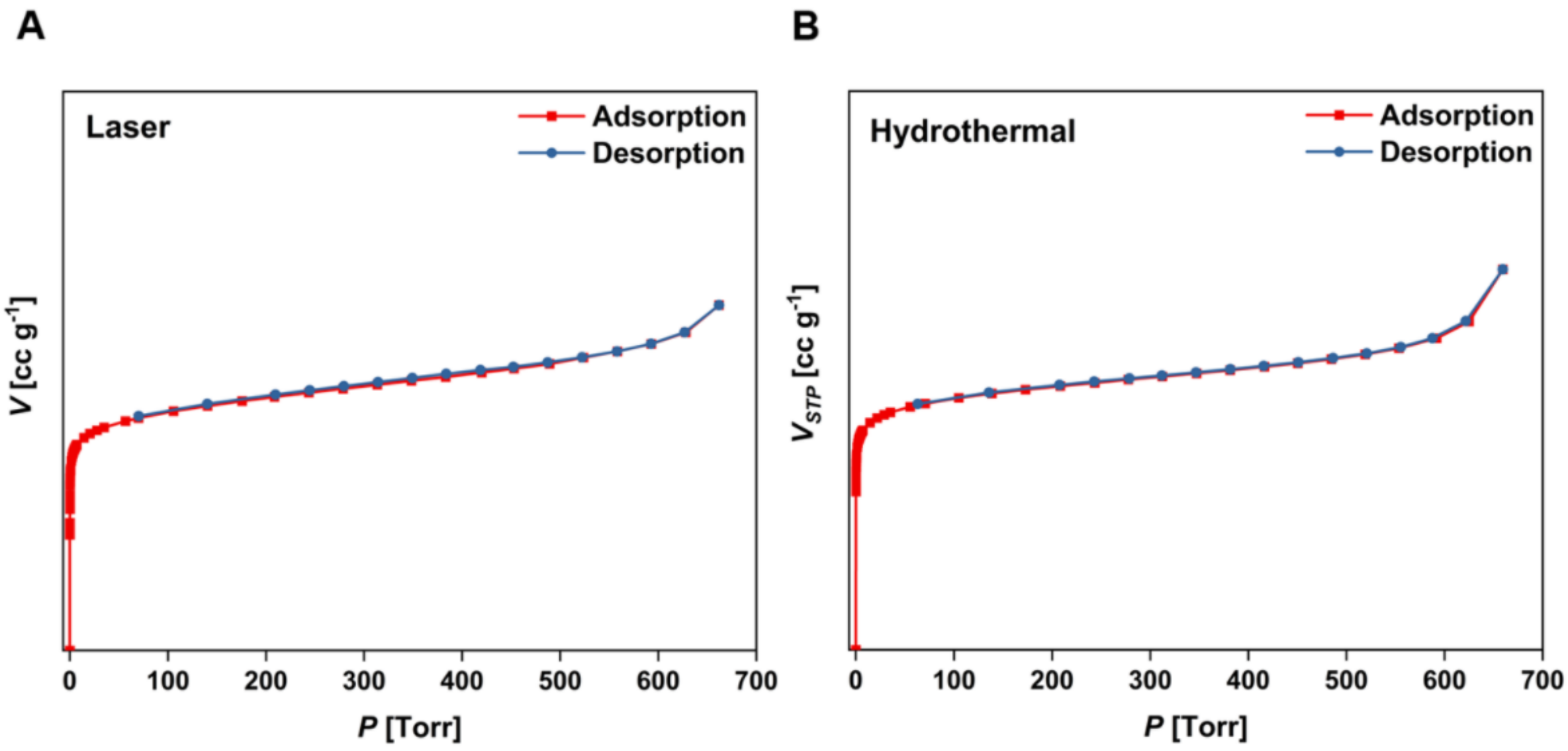

**Figure S13.** BET adsorption-desorption isotherms of TPA-Silicalite-1 zeolites (not calcined) synthesized with **(A)** laser and **(B)** hydrothermal methods. BET surface areas are calculated to be 335.5 $m^2\ g^{-1}$ for laser synthesized and 275.0 $m^2\ g^{-1}$ for hydrothermal synthesized crystals.

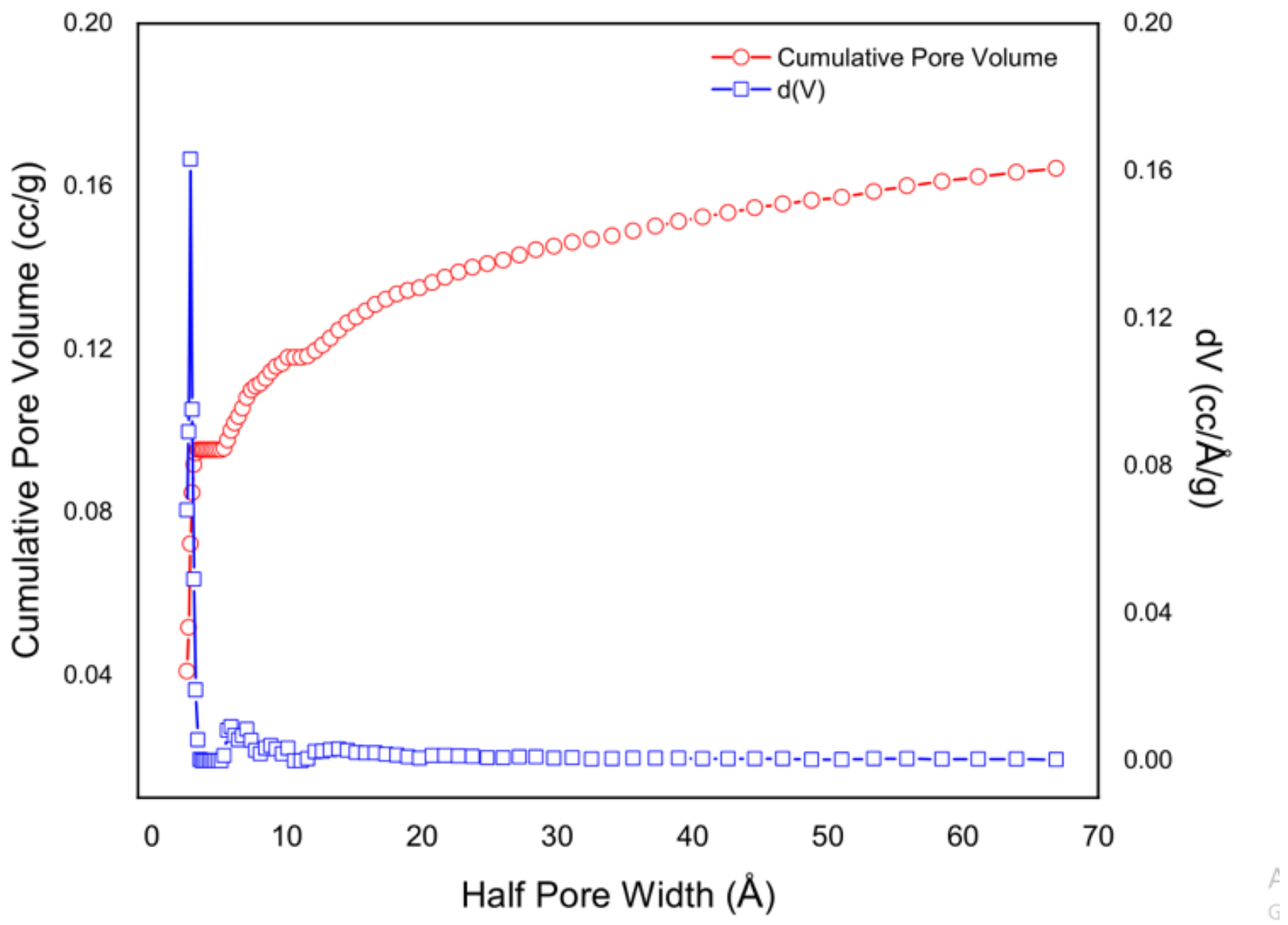

**Figure S14.** The cumulative pore volume and half-pore width distribution graphs for laser-synthesized TPA-Silicalite-1 zeolites. *dV* is the derivative of the cumulative pore volume with respect to pore width. The half-pore width, as determined by the DFT analysis, is 2.88 Å, and the micropore volume is 0.164 $cm^3\ g^{-1}$.

## ATR-FTIR peak deconvolution analyses

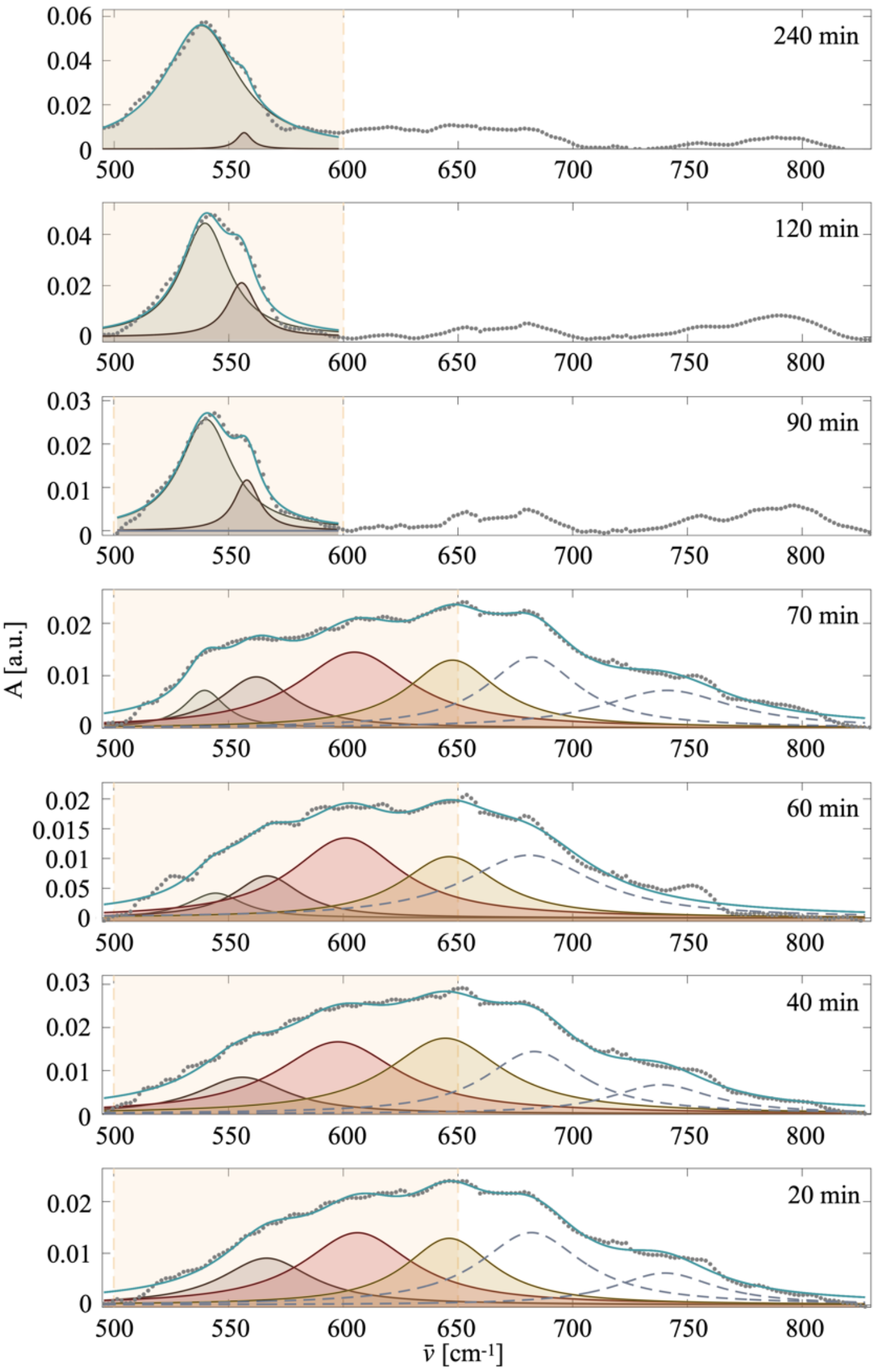

| Position | Height | FWHM | Area |
|---|---|---|---|
| 538 | 0.05584 | 39.13 | 2.648 |
| 556.7 | 0.007361 | 7.988 | 0.08757 |
| 539.6 | 0.04435 | 25.51 | 1.506 |
| 555.6 | 0.0211 | 15.09 | 0.4527 |
| 540.2 | 0.02567 | 27.7 | 0.9103 |
| 558 | 0.01168 | 13.34 | 0.2224 |
| 539.6 | 0.007196 | 21.75 | 0.2229 |
| 562.2 | 0.009766 | 38.15 | 0.5177 |
| 604.9 | 0.01447 | 60.11 | 1.187 |
| 647.6 | 0.0129 | 47.25 | 0.8665 |
| 682.4 | 0.01359 | 49.28 | 0.9476 |
| 741.1 | 0.007093 | 68.27 | 0.6339 |
| 544.3 | 0.004183 | 29.57 | 0.1721 |
| 566.9 | 0.007063 | 35.11 | 0.3497 |
| 601.2 | 0.01344 | 57.66 | 1.061 |
| 646.1 | 0.01031 | 50.56 | 0.7362 |
| 680.9 | 0.01055 | 69.04 | 0.9891 |
| 556.1 | 0.008461 | 47.14 | 0.5331 |
| 597.7 | 0.01667 | 64.59 | 1.446 |
| 644.5 | 0.01749 | 60.8 | 1.47 |
| 683.2 | 0.01442 | 51.92 | 1.053 |
| 738.5 | 0.006729 | 53.94 | 0.4947 |
| 566.6 | 0.009047 | 47.88 | 0.5878 |
| 606.2 | 0.01397 | 57.56 | 1.105 |
| 646.2 | 0.01289 | 46.69 | 0.8564 |
| 682 | 0.014 | 54.8 | 1.073 |
| 740.2 | 0.006171 | 53.15 | 0.4473 |

**Figure S15.** Peak deconvolution of the ATR-FTIR spectrum of laser-synthesized TPA-Silicalite-1 zeolites sampled at 20, 40, 60, 70, 90, 120, and 240 minutes of synthesis with position and height of each peak with their full-width at half maximum (FWHM) values and peak areas are color coded in the right panel.

**Table S1.** Supplementing Figure 2, the table presents the effect of laser parameters on the crystallinity and yield of laser-synthesized TPA-Silicalite-1 zeolite crystals.

| No | $V$ [µl] | $P_I$ [W] | $P_A$ [W] | $F_p$ [J cm$^{-2}$] | $E_p$ [µJ] | $t$ [h] | $f$ [kHz] | $E_{Tot}$ [MJ per $mol_{SiO_2}$] | $C$ [%][a),b)] | $Y$ [wt.%] [a)] |
|---|---|---|---|---|---|---|---|---|---|---|
| **T1** | 80 | 1 | 0.23 | 1.77 | 1.1 | 15 | 200 | 202.5 | - | - |
| **T2** | 80 | 1.7 | 0.38 | 1.77 | 1.1 | 9 | 333 | 202.5 | - | - |
| **T3** | 80 | 2.5 | 0.66 | 1.77 | 1.1 | 6 | 500 | 202.5 | 51 | 55.2 |
| **T4** | 80 | 5 | 1.13 | 1.77 | 1.1 | 3 | 1000 | 202.5 | 45.3 | 53.6 |
| **T5** | 80 | 2.5 | 0.56 | 0.88 | 0.6 | 6 | 1000 | 202.5 | - | - |
| **T6** | 80 | 5 | 1.13 | 3.54 | 2.3 | 3 | 500 | 202.5 | 87 | 69.7 |
| **T7** | 80 | 5 | 1.13 | 8.84 | 5.6 | 3 | 200 | 202.5 | 86 | 69.7 |

$V$ - Volume, $P_I$ - incident average laser power, $P_A$ - absorbed average laser power, $F_p$ - pulse fluence, $E_p$ - pulse energy, $t$ - time, $f$ - frequency, $E_{Tot}$ - total deposited energy, $C$ - crystallinity, $Y$ - yield, $T$ - trial.

**a)** The initial transparent color of the suspension did not change at the end of the reaction. Therefore, powder samples could not be collected for XRD analysis.

**b)** The reference sample used for calculating the crystallinity index is a laser-synthesized saturation sample, which was prepared under synthesis conditions of 5 W power, 200 kHz repetition rate, and a synthesis duration of 300 minutes. The crystallinity values are based on individual experimental results.

**Table S2.** The crystallinity index (%) and yield (wt.%) values, and average particle sizes of laser-synthesized TPA-Silicalite-1 zeolites using different molar formulas.

| | M1[a)] | | M2[b)] | | M3[c)] | |
|---|---|---|---|---|---|---|
| | Laser | Hydro. | Laser | Hydro. | Laser | Hydro. |
| *C* [%][d),e)] | 91 | 100 | 89 | 100 | 81 | 100 |
| *t* [h] | 3 | 48 | 3 | 30 | 3 | 30 |
| *Y* [wt.%] | 69.7 | 69.2 | 55.6 | 59.9 | 41.7 | 56.6 |
| $d_{Avg}$ ± SD [nm] | 323.6 ± 35.2 | 203.8 ± 50.4 | 175.8 ± 17.5 | 102.7 ± 19.1 | 167.1 ± 16.9 | 99.5 ± 16.0 |
| *CV* [%] | 10.9 | 24.7 | 10 | 18.6 | 10.1 | 16.1 |

*C* – crystallinity, *t* - time, *Y* - yield, $d_{Avg}$ - average particle size and *CV* - coefficient of variation.

a) M1 = 25 $SiO_2$: 9 TPAOH: 1450 $H_2O$: 100 EtOH.

b) M2 = 25 $SiO_2$: 9 TPAOH: 480 $H_2O$: 100 EtOH.

c) M3 = 25 $SiO_2$: 9 TPAOH: 450 $H_2O$: 100 EtOH.

d) The reference samples for calculating the crystallinity index are three different hydrothermal-synthesized saturation samples prepared using three different molar formulas to make each comparison consistent. The crystallinity values are based on individual experimental results.

e) The average particle sizes of the crystals were determined from SEM images using Image J software. SD indicates standard deviation. The sample size (N) for particle size distribution analyses was 298.

**Table S3.** The crystallinity index (%) and yield (wt.%) values of laser-synthesized TPA-Silicalite-1 zeolites for different reaction times.

| Molar Formula | t [min] | Color | Y [wt.%] | C [%][a)] |
|---|---|---|---|---|
| M1 | 70 | Light Milky | 9.3 | 48 |
| | 90 | Milky | 11.6 | 55 |
| | 120 | Opaque white | 46.5 | 62 |
| | 150 | Opaque white | 60.4 | 69 |
| | 180 | Opaque white | 69.7 | 78 |
| | 200 | Opaque white | 69.7 | 88 |
| | 220 | Opaque white | 74.4 | 96 |
| | 240 | Opaque white | 72.0 | 99 |
| | 260 | Opaque white | 74.4 | 98 |
| | 280 | Opaque white | 72.0 | 99 |
| | 300 | Opaque white | 76.7 | 100 |

*t* - time, *Y* - yield, and *C* - crystallinity.

**a)** The reference sample used for calculating the crystallinity index is a laser-synthesized saturation sample, which was prepared under synthesis conditions of 5 W power, 200 kHz repetition rate, and a synthesis duration of 300 minutes. The crystallinity values are averaged across five independent experiments.